\documentclass[a4paper]{article}
\usepackage[a4paper,top=2cm,left=2cm,right=2cm,bottom=2cm]{geometry}
\usepackage{amsmath}
\usepackage{amsthm}
\usepackage[latin1]{inputenc}
\usepackage{graphicx}
\usepackage{natbib}
\usepackage{algorithm}
\usepackage{setspace}
\usepackage{fancyhdr}
\usepackage{graphicx}
\usepackage{subfig}
\usepackage{rotating}
\usepackage{lscape}
\usepackage{amsmath}
\usepackage{amsthm}
\usepackage{algorithm}
\usepackage[noend]{algorithmic}
\usepackage{setspace}
\usepackage{graphicx,color}
\usepackage{hyperref}
\usepackage{enumitem}
\usepackage{amssymb}

\newcommand{\up}[1]{\raisebox{1.3ex}[0pt]{#1}}

\begin{document}
\large

\title{A matheuristic approach for the Pollution-Routing Problem}

\author{\textbf{Raphael Kramer, Anand Subramanian}\\
Departamento de Engenharia de Produção, Centro de Tecnologia \\ Universidade Federal da Paraíba \\
Campus I, Bloco G, Cidade Universitária, 58051-970, João Pessoa, PB \\
\{rkramer, anand\}@ct.ufpb.br
\and
\textbf{Thibaut Vidal} \\
Laboratory for Information and Decision Systems \\ Massachusetts Institute of Technology \\
77 Massachusetts Avenue, Cambridge, MA 02139, U.S. \\
vidalt@mit.edu
\and
\textbf{Lucídio dos Anjos Formiga Cabral} \\
Departamento de Computação Científica, Centro de Informática \\ Universidade Federal da Paraíba \\
Campus I, Cidade Universitária, 58051-900, João Pessoa, PB \\
lucidio@ci.ufpb.br
}
\date{}

\maketitle

\vspace{-0.5cm}
\begin{center}
Working Paper, UFPB -- April 2014 \\
\end{center}
\vspace{0.5cm}

\begin{abstract}

This paper deals with the Pollution-Routing Problem (PRP), a Vehicle Routing Problem (VRP) with environmental considerations, recently introduced in the literature by [Bekta\c{s} and Laporte (2011), Transport. Res. B-Meth. 45 (8), 1232-1250]. The objective is to minimize operational and environmental costs while respecting capacity constraints and service time windows. Costs are based on driver wages and fuel consumption, which depends on many factors, such as travel distance and vehicle load.  The vehicle speeds are considered as decision variables. They complement routing decisions, impacting the total cost, the travel time between locations, and thus the set of feasible routes. We propose a method which combines a local search-based metaheuristic with an integer programming approach over a set covering formulation and a recursive speed-optimization algorithm. This hybridization enables to integrate more tightly route and speed decisions.
Moreover, two other ``green'' VRP variants, the Fuel Consumption VRP (FCVRP) and the Energy Minimizing VRP (EMVRP), are addressed. The proposed method compares very favorably with previous algorithms from the literature and many new improved solutions are reported.

\end{abstract}

\onehalfspace

\section{Introduction}
\label{introduction}

The \emph{Vehicle Routing Problem} (VRP) and its variants have been the subject of considerable research efforts in the past year, mostly due to the growing number of additional constraints and objectives arising from real-world problems \citep{Vidal2013-survey}. Given the growing global concern about environmental issues, VRPs have recently started to incorporate ``green'' aspects such as pollution and alternative fuels, among others (see, e.g., the recent review of \citealt{Linetal2014} on VRPs with environmental issues).

Environmental costs due to greenhouse gas (GHG) emissions are not usually paid directly by the companies.
Nevertheless, some countries are developing emissions trading schemes (e.g., the European Union Emissions Trading System) to make the companies responsible for their environmental impacts.
The trend is that more and more countries will start to adopt emission-reducing actions, enhancing the importance of VRPs with environmental considerations. Furthermore, CO$_2$ is the most prominent transportation GHG, and its emission is closely related to fuel consumption \citep{ICF2006}.


In this work we turn our attention to the Pollution-Routing Problem (PRP), recently introduced by \cite{BektasLaporte2011}. The PRP is $\mathcal{NP}$-hard since it includes the VRP with Time Windows (VRPTW) as a particular case, and most current exact procedures cannot address problems of practical sizes.
A recent metaheuristic \citep{DemirBL12} has thus been proposed for the problem. However, although routing and speed decisions are tightly related in the problem formulation, these variables are optimized sequentially in past methods by first solving a routing problem with some fixed initial speeds, and then performing speed optimization as a post-optimization procedure. 

This paper introduces a novel hybrid algorithm that integrates a Speed Optimization Algorithm (SOA) and a Set Partitioning (SP) approach into a multi-start Iterated Local Search (ILS) framework.
The method attempts to tightly integrate routing and speed decisions by performing a quick sequence of route and speed optimizations. Furthermore, it has been recently shown by \citet{Subramanianetal2012, Subramanianetal2013} that ILS coupled with SP yields competitive results for various VRPs that do not consider time-window constraints. The algorithm presented in this work generalizes these method by efficiently handling time-window constraints and considering some infeasible solutions during shaking. Move evaluations are performed in amortized $\mathcal{O}(1)$ time by extending the scheme developed by \citet{Vidaletal2013}.
The high performance of the method is demonstrated by extensive computational experiments on the existing instances for the PRP, characterized by large time windows, as well as newer difficult instances with tighter time windows. Moreover, two particular cases of the PRP are studied: the Fuel Consumption Vehicle Routing Problem (FCVRP) \citep{Xiaoetal2012} and the Energy Minimizing Vehicle Routing Problem (EMVRP) \citep{Karaetal2007}. Again, the proposed method outperforms existing methods from the literature.

The remainder of this paper is organized as follows. Section \ref{RelatedWork} presents some works that integrate VRPs with environmental aspects. Section \ref{ProblemDescription} formally defines the PRP, EMVRP and FCVRP. Section \ref{Algorithm} describes the proposed approach. Computational results are provided in Section \ref{Results}, while Section \ref{Conclusions} concludes.

\section{Related works and challenges}
\label{RelatedWork}

\citet{Palmer2007} was the first to incorporate environmental issues to the VRP. Different from the previous works that estimated the environmental costs based on the total duration or distance of the routes, the author considered other issues such as road topography, congestion and vehicle speeds to generate a CO$_2$ emissions matrix. Experiments suggested that the CO$_2$ minimization model compared with the distance-minimizing and duration-minimizing models led to a CO$_2$ reduction of $5.20\%$ and $5.02\%$, respectively, on average. 

Later on,
\citet{Karaetal2007} proposed a mathematical formulation for the so-called \emph{Energy Minimizing Vehicle Routing Problem} (EMVRP), which aims at minimizing the sum of the product between load and distance for each arc.
Similar approaches, i.e, those that make use of the vehicle load to minimize the fuel consumption or CO$_2$ emissions, were presented by \citet{PengWang2009}, \citet{Scottetal2010}, \citet{Ubedaetal2011} and \citet{Xiaoetal2012}. The latter introduced the \emph{Fuel Consumption Vehicle Routing Problem} (FCVRP). \citet{Karaetal2008} presented a VRP with cumulative costs, which generalizes the EMVRP as well as the \emph{Minimum Latency Problem} and the \emph{m-Traveling Repairman Problem}. \citet{Kopferetal2013} took the load into account for CO$_2$ emissions evaluations in the presence of heterogeneous vehicle types.


Minimizing the fuel consumption considering only the load and distance can be insufficient since the travel speed plays a major role. This speed is directly affected by the road congestion. In view of this, \citet{Kuo2010} proposed a model for minimizing the total fuel consumption where the speeds are time-dependent and the load is used to estimate the cost. A Simulated Annealing algorithm was implemented to solve a set of instances from the Solomon benchmark \citep{Solomon1987}. Later, \citet{KuoWang2011} devised a Tabu Search algorithm for the same problem. Other time-dependent problems with emission minimization can be found in \citet{Figliozzi2011}, \citet{SaberiVerbas2012} and \citet{Jabalietal2012}, and many other references relevant to green logistics are mentioned in the surveys of \citet{Dekkeretal2012}, \citet{Salimifardetal2012}, \citet{Linetal2014} and \citet{Demiretal2014b}.

\citet{BektasLaporte2011} proposed the \emph{Pollution-Routing Problem} (PRP), which seeks to minimize both operational and environmental costs, taking into account the customers time-window constraints. The total travel distance, the amount of load carried per distance unit, the vehicle speeds and the duration of the routes are the main cost components.
Three different variants were also presented, considering either distance, weighted load and energy minimization.
The authors performed an extensive experimental analysisto capture the trade-off between each variation, as well as the effect of the speed and time-window constraints on the distance, energy and costs.

The PRP was addressed with a two-phase heuristic in \citet{DemirBL12}. In the first phase, the VRPTW is solved by means of an Adaptive Large Neighborhood Search (ALNS), including five insertion operators and twelve removal operators. 
In a second phase, vehicle speeds are optimized using a recursive algorithm. Computational experiments were carried out for instances with up to 200 customers. Other recent developments consider generalizations of the problem. A bi-objective variant considering fuel and driving time minimization is presented in \citet{Demiretal2014}, and \citet{Franceschettietal2013} consider the time-dependent PRP.

It should be noted that several aspects of the PRP are conflicting. Higher speeds imply routes with shorter durations, but at the same time result in a larger amount of emissions and vice-versa.
Hence, to reduce pollution, speed on arcs may be decreased to be closer to the speed which minimizes emissions. Yet with lower speed, the set of feasible VRP routes may be empty or drastically smaller. As a consequence, an optimal solution of the reduced speed VRP can have longer distance and even more emissions in some cases.
One main goal of our study was to better integrate route and speed decisions in order to find quickly a suitable balance between these antagonist aspects.

\section{Problem description}
\label{ProblemDescription}

The PRP can be defined as follows. Let $\mathcal{G}=(\mathcal{V},\mathcal{A})$ be a complete and directed graph with a set $\mathcal{V}=\{0,1,2,\dots,n\}$ of vertices and a set $\mathcal{A}=\{(i,j) \in \mathcal{V}^2, i \neq j\}$ of arcs. Vertex $0$ represents the depot where a fleet of $m$ identical vehicles with capacity $Q$ is based. Vertices $\mathcal{V}-\{0\}$ correspond to customers, characterized by a non-negative demand $q_i$ for a single product, a service time $\tau_i$ and a specified time-window interval $[a_i, b_i]$ for service. We assume that $q_0 = 0$ and  $\tau_0 = 0$. Each arc $(i,j) \in \mathcal{A}$ represents a travel possibility from node $i$ to $j$ for a distance $d_{ij}$.

A particularity of the PRP is that the speed $v_{ij}$ on each $(i,j)$ is itself a decision variable, valued between $[v_\textsc{min},v_\textsc{max}]$. Indeed, each vehicle emits on a certain amount of GHG which depends of weight and speed, among other factors. The PRP aims to find a speed matrix $(\mathbf{v})_{ij}$ and a set of routes $\mathbf{R}$ (such that $|\mathbf{R}| \leq m$)  to serve all customers while minimizing environmental and operational costs. Each route $\boldsymbol\sigma \in \mathbf{R}$, $\boldsymbol\sigma = (\sigma_1,\dots,\sigma_{|\sigma|}) $ starts and ends at the depot, i.e., $\sigma_1 = 0$ and $\sigma_{|\sigma|} = 0$, the total demand on each route should not exceed the vehicle capacity, and every customer must be visited within its time window.

Equation (\ref{routeLoad}) defines for a route $\boldsymbol\sigma$ the vehicle load $f_{\sigma_i \sigma_{i+1}}$ when traveling on arc $(\sigma_i, \sigma_{i+1})$, and Equation (\ref{routeArrivalTimes}) defines recursively the arrival time $t_i$ to customers, knowing that each route starts at time zero, that it is allowed to arrive early at a customer location and wait the start of its time window $a_i$, but that a late arrival is not permitted.
\begin{equation}
\label{routeLoad}
f_{\sigma_i \sigma_{i+1}} = \sum_{k=i+1}^{|\sigma|} q_{\sigma_k}, \hspace*{0.5cm} i  = 1,\dots,|\sigma|-1
\end{equation}
\begin{equation}
\label{routeArrivalTimes}
\left\{
\begin{aligned}
t_{\sigma_1} &= 0 \\
t_{\sigma_i} &= \max \left\{ a_{\sigma_{i-1}}, t_{\sigma_{i-1}}\right\} + \tau_{\sigma_{i-1}} +  \frac{d_{\sigma_{i-1}\sigma_{i}}}{v_{\sigma_{i-1}\sigma_{i}}},   \hspace*{0.5cm} i  = 2,\dots,|\sigma| \\
\end{aligned}
\right.
\end{equation}

The PRP objective is based on the assumption that CO$_2$ emissions are approximately proportional to fuel consumption. Using the comprehensive emissions model of \citet{Barthetal2005}, \citet{ScoraBarth2006} and \citet{BarthBoriboonsomsin2008}, \cite{BektasLaporte2011}  obtained the consumption profile $F^\textsc{f}_{\sigma_{i}\sigma_{i+1}}(v_{\sigma_{i}\sigma_{i+1}})$ of Equation (\ref{fc1}) for an arc $(\sigma_{i},\sigma_{i+1})$ at speed $v_{\sigma_{i}\sigma_{i+1}}$. Parameters $w_1,w_2,w_3, w_4$ are based on fuel properties, vehicle and network characteristics. Their meaning and value are discussed in Appendix 1, based on \cite{BektasLaporte2011} and \citet{DemirBL12}. Finally, the overall objective of the PRP, given in Equation (\ref{PRPObj}), considers both the fuel consumption with a cost of $\omega_{\textsc{fc}}$ pounds (\textsterling) per liter, and the driving costs at a rate of $\omega_{\textsc{fd}}$ per unit of time.
\begin{align}
F^\textsc{f}_{\sigma_{i}\sigma_{i+1}}(v_{\sigma_{i}\sigma_{i+1}}) &= d_{\sigma_{i}\sigma_{i+1}}  \left( \frac{w_1}{v_{\sigma_{i}\sigma_{i+1}}} + w_2 + w_3 f_{\sigma_{i}\sigma_{i+1}} + w_4v_{\sigma_{i}\sigma_{i+1}}^{2} \right)
 \label{fc1} \\
 Z_{\textsc{prp}}(\mathbf{R},\mathbf{v}) &= 
\sum_{\boldsymbol\sigma \in \mathbf{R}}  \left( \omega_{\textsc{fc}}  \sum_{i=1}^{|\sigma|-1} F^\textsc{f}_{\sigma_{i}\sigma_{i+1}}(v_{\sigma_{i}\sigma_{i+1}}) + \omega_{\textsc{fd}} \ t_{\sigma_{|\sigma|}} \right)
  \label{PRPObj}
\end{align}

The fuel consumption $F^\textsc{f}_{\sigma_{i}\sigma_{i+1}}(v_{\sigma_{i}\sigma_{i+1}})$ is convex. The minimum $v^*_\textsc{f}$ on this function, i.e., the speed value that minimizes fuel costs, is given in Equation~(\ref{minF}).
\begin{equation}
\label{minF}
\frac{d F^\textsc{f}_{\sigma_{i}\sigma_{i+1}}}{d v_{\sigma_i\sigma_{i+1}}} (v^*_\textsc{f}) = 0  
\Leftrightarrow  v^*_\textsc{f} = \left( \frac{w_1}{2 w_4} \right)^{1/3}
 \end{equation}

Similarly, for any arc $(\sigma_i,\sigma_{i+1})$, assuming that there is no waiting time in the route after $\sigma_{i}$, the travel cost $F^\textsc{fd}_{\sigma_{i}\sigma_{i+1}}(v_{\sigma_{i}\sigma_{i+1}})$ including driver wages is given in Equation (\ref{fueldriver}). The speed value that minimizes fuel and driver costs is then expressed in Equation (\ref{minFD}). Both values,  $v^*_\textsc{f}$ and $v^*_\textsc{fd}$, are independent of the arc under consideration.
\begin{equation}
\label{fueldriver}
F^\textsc{fd}_{\sigma_{i}\sigma_{i+1}}(v_{\sigma_{i}\sigma_{i+1}}) = \omega_{\textsc{fc}} d_{\sigma_{i}\sigma_{i+1}}  \left( \frac{w_1}{v_{\sigma_{i}\sigma_{i+1}}} + w_2 + w_3 f_{\sigma_{i}\sigma_{i+1}} + w_4v_{\sigma_{i}\sigma_{i+1}}^{2} \right) + \omega_{\textsc{fd}} \frac{d_{\sigma_i\sigma_{i+1}}}{v_{\sigma_i\sigma_{i+1}}}
\end{equation}
\begin{equation}
\label{minFD}
\frac{d F^\textsc{fd}_{\sigma_{i}\sigma_{i+1}} }{d v_{\sigma_i\sigma_{i+1}}} (v^*_\textsc{fd}) = 0 
\Leftrightarrow  v^*_\textsc{fd} = \left( \frac{\frac{\omega_\textsc{fd}}{\omega_\textsc{fc}}  + w_1}{2 w_4} \right)^{1/3}
 \end{equation}

The two other problems considered in this work, EMVRP and FCVRP, do not take into account time-window constraints and speed decisions. In practice, these problems can be seen as CVRPs with green-oriented objective functions.
The objective of the EMVRP is based on a simplified emission model, which takes solely into account the distance and the load$\times$distance factor \citep{Karaetal2007}. It is expressed in Equation (\ref{EMVRPObj}), where $\omega$ represents the weight of a vehicle without cargo.
\begin{equation}
Z_{\textsc{emvrp}}(\mathbf{R},\mathbf{v}) = \sum_{\boldsymbol\sigma \in \mathbf{R}}    \sum_{i=1}^{|\sigma|-1}   d_{\sigma_{i}\sigma_{i+1}}  (\omega + f_{\sigma_{i}\sigma_{i+1}})
 \label{EMVRPObj} 
\end{equation}

Finally, the FCVRP is based on a linear function of fuel consumption per distance, which considers the vehicle's no-load/curb weight, the load carried, and the fixed cost of the vehicle. The objective is expressed in Equation (\ref{FCVRPObj}), where $h$ is the vehicle fixed cost, $\rho^*$ and $\rho_0$ are the fuel consumption rate of the vehicle without any load and fully loaded, respectively.
\begin{equation}
Z_{\textsc{fcvrp}}(\mathbf{R},\mathbf{v}) =  \sum_{\boldsymbol\sigma \in \mathbf{R}}
\left( 
h + \omega_{\textsc{fc}} \sum_{i=1}^{|\sigma|-1}   
\left( 
 d_{\sigma_{i}\sigma_{i+1}}  ( \rho_0 +  \frac{\rho^* - \rho_0}{Q}  f_{\sigma_{i}\sigma_{i+1}} )
\right)
\right)
 \label{FCVRPObj}
\end{equation}

Using the same values of \citet{Xiaoetal2012}, with $h = 0$, $\omega_{\textsc{fc}} = 1$, $\rho^* = 2$, and $\rho_0 = 1$, the objective can be reformulated as in Equation (\ref{FCVRPObj2}). It is noteworthy that, as in the case of the EMVRP, the FCVRP involves a linear combination of the distance and the load$\times$distance factor.
\begin{equation}
Z_{\textsc{fcvrp}}(\mathbf{R},\mathbf{v}) =  
\sum_{\boldsymbol\sigma \in \mathbf{R}} \sum_{i=1}^{|\sigma|-1}   d_{\sigma_{i}\sigma_{i+1}}  \left( 1 + \frac{1}{Q}  f_{\sigma_{i}\sigma_{i+1}}  \right)
 \label{FCVRPObj2}
\end{equation}

\section{The proposed ILS-SP-SOA matheuristic}
\label{Algorithm}

The proposed algorithm, called ILS-SP-SOA, combines an iterated local search with speed optimization procedures and integer programming optimization over a set partitioning formulation. As a blend of metaheuristic and exact procedures, this category of method is usually called \emph{matheuristic} \citep{ManSV2009}.

In ILS-SP-SOA, the initial solution construction and local search, as well as three perturbation procedures aim at minimizing the cost by considering routing decisions without changing the speeds. This sub-problem can be seen as a VRPTW with the objective of Equation (\ref{PRPObj}). We demonstrate in Section \ref{sec:LocalSearch} how local-search moves can be evaluated in $O(1)$ time for this setting.
The search is complemented by a recursive speed-optimization procedure, described in Section \ref{sec:SOA}, which is applied on each local minimum of the ``routing'' local search. New speed decisions are included in a dynamic speed matrix, which is in turn used in the subsequent VRPTW sub-problems.
Finally, the routes associated to local minimums are stored in a pool, and used by an integer optimization procedure over a set partitioning formulation to generate possible better solutions composed of a different recombination of routes. In this process, each route may be associated to a different speed matrix, and thus this exact procedure adds an additional level of integration between route and speed decisions.

The outline of ILS-SP-SOA is presented in Algorithm \ref{ILS-SP-SOA}.
The method performs $n_\textsc{r}$ restarts (Lines \ref{ILS-SP-SOA:beg}-\ref{ILS-SP-SOA:end}) of a hybrid procedure combining ILS, SOA and SP. 
At each restart, the speed matrix $\mathbf{v}$ is first initialized with the maximum speed, $v_{ij} = v_\textsc{max}$ for all $(i,j) \in \mathcal{A}$. A solution $S$ is obtained by applying local search and SOA on an initial solution (Line~\ref{ILS-SP-SOA:InitSolLSSOA}). This initial solution is generated using the modified cheapest insertion heuristic of \citet{Pennaetal2013}, considering the PRP objective and the time-window relaxation of Section \ref{sec:LocalSearch}.
The speed of arcs associated to $S$ is then updated in the matrix $\mathbf{v}$.

\begin{algorithm}[htb]
\caption{ILS-SP-SOA($n_\textsc{r}$, $n_\textsc{ils}$, $n_\textsc{sp}$, $n_\textsc{pool}$, $T_\textsc{mip}$, seed)}
\label{ILS-SP-SOA}
\begin{algorithmic}[1]
\algsetup{indent=2.5em}
\STATE $S_\textsc{best-all} \leftarrow \emptyset$; $f(S_\textsc{best-all}) \leftarrow \infty$; \COMMENT{$S_\textsc{best-all}$ is the overall best solution}
\STATE  $P_\textsc{perm}$ $\leftarrow$ $\emptyset$; $i_\textsc{r} = 0$; \COMMENT{ $P_\textsc{perm}$ is the pool of permanent routes in the SP}
\WHILE {$i_\textsc{r} < n_\textsc{r}$} \label{ILS-SP-SOA:beg}
     \STATE $i_\textsc{r}$ $\leftarrow$ $i_\textsc{r} + 1$;
    \STATE $S_\textsc{best} \leftarrow \emptyset$; $f(S_\textsc{best}) \leftarrow \infty$; \COMMENT{$S_\textsc{best}$ is the best solution of the restart phase}
  \STATE $P_\textsc{temp}$ $\leftarrow$ $\emptyset$; $i_\textsc{ils}$ $\leftarrow$ 0; \COMMENT{ $P_\textsc{perm}$ is the pool of temporary routes in the SP}
    \STATE $\mathbf{v} \leftarrow$ InitializeSpeedMatrix($v_\textsc{max}$); \label{ILS-SP-SOA:initspeed}
    \STATE $S$ $\leftarrow$ SpeedOptimization(LocalSearch(GenInitSol(seed))); \label{ILS-SP-SOA:InitSolLSSOA}
    \STATE $\mathbf{v} \leftarrow$ UpdateSpeedMatrix($S$);  \label{ILS-SP-SOA:updatespeed1}
    \WHILE {$i_\textsc{ils} < n_\textsc{ils}$} \label{ILS-SP-SOA:ILS-beg}
      \STATE $i_\textsc{ils}$ $\leftarrow$ $i_\textsc{ils} + 1$;
        \STATE $S$ $\leftarrow$ SpeedOptimization(LocalSearch(Perturbation($S_\textsc{best}$, seed))); \label{ILS-SP-SOA:InitSolLSPert}
 \STATE $\mathbf{v} \leftarrow$ updateSpeedMatrix($S$);  \label{ILS-SP-SOA:updatespeed2}
    \STATE $P_\textsc{temp}$ $\leftarrow$ $P_\textsc{temp} \cup (\text{Feasible routes of } S$); \label{ILS-SP-SOA:TempRoutePool}
    \IF {$f(S) < f(S_\textsc{best}$)}
      \STATE $S_\textsc{best}$ $\leftarrow$ $S$; $i_\textsc{ils}$ $\leftarrow$ 0;
    \ENDIF \label{ILS-SP-SOA:ILS-end}
    \IF {$i_\textsc{ils} \geq n_\textsc{ils}/2$} \label{ILS-SP-SOA:reinitspeed1}
      \STATE $\mathbf{v} \leftarrow$ ReinitializeSpeedMatrix($v_\textsc{max}$, $S_\textsc{best}$); \label{ILS-SP-SOA:reinitspeed2}
    \ENDIF
 \ENDWHILE
 \IF {($n \leq n_\textsc{sp}$ \textbf{and} $i_\textsc{r} = n_\textsc{r} - 1$) \textbf{or} $n > n_\textsc{sp}$} \label{ILS-SP-SOA:SP-beg}
  \STATE $S_\textsc{best}\leftarrow$ MIPSolver($S_\textsc{best}$, $P_\textsc{temp}$, $P_\textsc{perm}$, $n_\textsc{ils}$, $T_\textsc{mip}$); \label{ILS-SP-SOA:SP-MIP}
 \ENDIF
    \IF {$f(S_\textsc{best}) < f(S_\textsc{best-all})$}
        \STATE $S_\textsc{best-all} \leftarrow S_\textsc{best}$;
    \ENDIF
 \STATE $P_\textsc{perm}$ $\leftarrow$ $P_\textsc{perm} \cup (\text{Feasible routes of } S_\textsc{best}$); \label{ILS-SP-SOA:end} \label{ILS-SP-SOA:Perm}
    \IF {number of consecutive restarts $\geq n_\textsc{pool}$} \label{ILS-SP-SOA:Clean1}
  \STATE $P_\textsc{temp}$ $\leftarrow$ $\emptyset$;\label{ILS-SP-SOA:Clean2}
    \ENDIF
\ENDWHILE
\RETURN{$S_\textsc{best-all}$} \label{ILS-SP-SOA:best}
\end{algorithmic}
\end{algorithm}

The ILS is then run until $n_\textsc{ils}$ successive iterations of local search and perturbation are reached without improvement.
At each iteration (Lines \ref{ILS-SP-SOA:ILS-beg}-\ref{ILS-SP-SOA:reinitspeed2})
the local optimal solution is modified by one of the three perturbation mechanisms, selected at random with different probabilities as described in Section \ref{sec:Perturbation}.
Time-window constraints are relaxed when performing a perturbation move. This modified solution is possibly improved by applying local search and SOA (Line \ref{ILS-SP-SOA:InitSolLSPert}), and the speed matrix is updated (Line \ref{ILS-SP-SOA:updatespeed2}). A temporary pool of routes is updated during the ILS loop by adding routes associated to local optimal solutions (Line \ref{ILS-SP-SOA:TempRoutePool}). If the number of ILS iterations without improvement is equal to $n$, the perturbation \emph{Change Speeds} is applied, as explained later in Subsection \ref{sec:Perturbation}.
Finally, the speed matrix $\mathbf{v}$ is reinitialized when the number of ILS iterations without improvement is greater than $n_\textsc{ils}/2$. (Lines \ref{ILS-SP-SOA:reinitspeed1}-\ref{ILS-SP-SOA:reinitspeed2}).

If the size of the instance is smaller than a given parameter $n_\textsc{sp}$, then the SP method is called only after the last restart phase; otherwise, it is called after every restart. The SP method attempts to create a new solution from routes of a temporary pool of routes $P_\textsc{temp}$ derived from local optimums of the local search, which is cleared after each $n_\textsc{pool}$ restarts (Lines \ref{ILS-SP-SOA:Clean1}-\ref{ILS-SP-SOA:Clean2}), and from a permanent pool of routes $P_\textsc{perm}$ which contains the routes associated to the best solutions $S_ \textsc{best}$ of each restart phase (Line \ref{ILS-SP-SOA:Perm}). The SP problems are solved using a Mixed Integer Programming (MIP) solver which calls the ILS scheme every time a new incumbent solution is found (Line \ref{ILS-SP-SOA:SP-MIP}).
This collaborative approach, described in \cite{Subramanianetal2013}, not only potentially speeds up the solver runtime, but also may contribute to find better solutions than the best possible combination of routes in the pools. If this happens, the pool $P_\textsc{temp}$ is also updated.
The MIP solver is run several consecutive times until no improvement is found over $S_\textsc{best}$, and a time limit $T_\textsc{mip}$ is imposed to each MIP execution to avoid any unpredictable excessive CPU time. Finally, the algorithm returns the best solution found among all restarts (Line \ref{ILS-SP-SOA:best}).

\subsection{Local Search with efficient move evaluation}
\label{sec:LocalSearch}

The local search procedure is based on the Randomized Variable Neighborhood
Descent (RVND) of \citet{Subramanianetal2010}. It relies on five inter-route
neighborhoods, namely: \emph{Shift(1,0)}, \emph{Shift(2,0)}, \emph{Swap(1,1)}, \emph{Swap(2,2)}, \emph{2-opt$^*$} and five intra-route neighborhoods, namely: \emph{Reinsertion}, \emph{Or-opt2}, \emph{Or-opt3}, \emph{Exchange} and \emph{2-opt}. The neighborhood structures based on
\emph{Shift} operators move one or more consecutive customers from a route to
another one. Those based on \emph{Swap} consists of interchanging one or more
customers from one route with one or more customers from another.
\emph{Reinsertion} and \emph{Or-opt} follow the idea of \emph{Shift}, but
involving customers of a single route. Exchange is the intra-route version of
\emph{Swap(1,1)}. Finally, \emph{2-opt} is the classical Traveling Salesman
Problem (TSP) intra-route operator, whereas \emph{2-opt$^*$} is the
inter-route version of \emph{2-opt}. A detailed description of these
neighborhoods, as well as the Auxiliary Data Structures (ADSs) used to enhance
the performance the local search can be found in \cite{Subramanian2012}, \citet{Pennaetal2013} and \citet{Vidaletal2013}. 

A local search for the VRPTW may have difficulties to generate feasible solutions, possibly compromising the convergence towards good solutions. To circumvent this issue, penalized infeasible solutions w.r.t. time windows are considered, and the objective includes ``time-warp'' penalties as in 
\citep{Vidaletal2013}.
To compute the cost of new routes generated by the local search in amortized constant time, other ADSs based on subsequences concatenation were implemented. For any subsequence $\sigma$ the algorithm stores and maintains: 
\begin{itemize}
\vspace{-.25cm} \item minimum duration $T(\sigma)$,
\vspace{-.25cm} \item minimum time-warp use $TW(\sigma)$,
\vspace{-.25cm} \item earliest $E(\sigma)$ and latest visit $L(\sigma)$ to the first vertex allowing a schedule with minimum duration and minimum time-warp use,
\vspace{-.25cm} \item cumulated load $Q(\sigma)$,
\vspace{-.25cm} \item distance $D(\sigma)$,
\vspace{-.25cm} \item travel time $TT(\sigma)$,
\vspace{-.25cm} \item load $\times$ distance $QD(\sigma)$,
\vspace{-.25cm} \item and speed$^2$ $\times$ distance $SSD(\sigma)$.
\end{itemize}

For a subsequence $\bar{\sigma}$ involving a single customer, $i$, these ADSs are computed as follows: $T(\bar{\sigma}) = \tau_i$; $TW(\bar{\sigma}) = 0$; $E(\bar{\sigma}) = a_i$; $L(\bar{\sigma}) = b_i$; $Q(\bar{\sigma}) = q_i$; $D(\bar{\sigma}) = 0$; $TT(\bar{\sigma}) = 0$; $QD(\bar{\sigma}) = 0$; $SSD(\bar{\sigma}) = 0$. If the first node in the sequence is a depot, e.g., if $\bar{\sigma} = 0$, then $E(\bar{\sigma}) = L(\bar{\sigma}) = 0$. This prevents a delayed departure. The following equations enable then to derive ADSs for larger subsequences obtained by concatenation $\oplus$:
\begin{align}
 \Delta &= T(\sigma) - TW(\sigma) + \delta_{\sigma_{|\sigma|} \sigma'_1}\label{structures::delta}\\ 
 \Delta WT &= \max\{E(\sigma') - \Delta - L(\sigma), 0\}\\
 \Delta TW &= \max\{E(\sigma) + \Delta - L(\sigma'), 0\}\\
 T(\sigma \oplus \sigma') &= T(\sigma) + T(\sigma') + \delta_{\sigma_{|\sigma|} \sigma'_1} + \Delta WT\\
 TW(\sigma \oplus \sigma') &= TW(\sigma) + TW(\sigma') + \Delta TW\\
 E(\sigma \oplus \sigma') &= \max\{E(\sigma') - \Delta, E(\sigma)\} - \Delta WT\\
 L(\sigma \oplus \sigma') &= \min\{L(\sigma') - \Delta, L(\sigma)\} + \Delta TW\\
 Q(\sigma \oplus \sigma') &= Q(\sigma) + Q(\sigma')\\
 D(\sigma \oplus \sigma') &= D(\sigma) + D(\sigma') + d_{\sigma_{|\sigma|} \sigma'_1} \\
 TT(\sigma \oplus \sigma') &= TT(\sigma) + TT(\sigma') + \delta_{\sigma_{|\sigma|} \sigma'_1}\\
 QD(\sigma \oplus \sigma') &= QD(\sigma) + QD(\sigma') + Q(\sigma')(D(\sigma) + d_{\sigma_{|\sigma|} \sigma'_1})\\
 SSD(\sigma \oplus \sigma') &= SSD(\sigma) + SSD(\sigma') + v_{\sigma_{|\sigma|} \sigma'_1}^2 d_{\sigma_{|\sigma|} \sigma'_1} \label{structures::SSD}
\end{align}
The penalized cost of a route $\sigma$ can be derived from these structures as shown in Equation (\ref{computePRPCost}) where $\omega_\textsc{tw}$ is the penalty for one unit of time warp.
\begin{align}
Z(\sigma) & = \omega_{\textsc{fc}} \left( w_1 TT(\sigma) + w_2 D(\sigma) + w_3 QD(\sigma) + w_4 SSD(\sigma) \right)  \nonumber \\ &+ \omega_{\textsc{fd}} T(\sigma) + \omega_\textsc{tw} TW(\sigma)
\label{computePRPCost}
\end{align}

\subsection{Perturbation Mechanisms}
\label{sec:Perturbation}

The three mechanisms applied during the perturbation phase are described as follows.

\begin{itemize}
 \item \emph{Shift to End} --- Shifts one customer from one route to the end of other route, randomly with uniform distribution.
 \item \emph{Merge routes} --- The two routes with the smallest accumulated load are merged if the vehicle capacity is not exceeded in the process. Given two routes $\sigma$ and $\sigma'$, the new merged route will be $\hat{\sigma} = (0, \sigma_2, \dots, \sigma_{n-1}, \sigma'_2, \dots, \sigma'_{n-1}, 0)$.
 \item \emph{Change Speeds} ---  This perturbation modifies the speed matrix by changing the speeds associated to arcs of one random route in the current best solution $S_\textsc{best}$ with one random speed from the set $\{v^{*}_\textsc{f}, v^{*}_\textsc{fd}, v_\textsc{max}\}$, with uniform distribution.
\end{itemize}

During the perturbation phase, \emph{Shift to End} and \emph{Merge routes} are randomly selected with different probabilities, $90\%$ and $10\%$, respectively. Any perturbation leading to an infeasible solution w.r.t. capacity constraints is undone and a new perturbation is attempted.
Finally, the additional \emph{Change Speeds} perturbation is applied after $n$ consecutive ILS iterations without improvement, e.g., when $i_\textsc{ils} = n$.

\subsection{Speed Optimization Algorithm}
\label{sec:SOA}

Speeds have a direct impact on customers' arrival time, i.e., in meeting their time windows. When routes are fixed, the PRP leads to a speed optimization problem which consists of finding the optimal speeds for each arc while respecting customers' time windows. This problem is solved by a recursive algorithm (Algorithm \ref{ArrivalTimes}). This method is an adaptation of the RSA procedure of \cite{Norstadetal2011} and \cite{Hvattumetal2013}, also considering drivers wages, possible waiting times, and a non-fixed arrival time at the last customer.

Algorithm \ref{ArrivalTimes} is applied on the complete route, by setting $s=1$ and $e=|\sigma|$. Let $t'_{\sigma_{i}}$ be the time to start the service of customer $\sigma_{i}$. Early and late arrivals are not considered in the course of the RSA procedure such that $t'_{\sigma_{|\sigma|}}$ is also the arrival time at customer $\sigma_{i}$.
The algorithm first computes the arrival time  $t'_{\sigma_{|\sigma|}}$ at the last customer when traveling at speed $v^{*}_\textsc{fd} $ (minimizing fuel consumption plus driver costs). If this time $t'_{\sigma_{|\sigma|}}$ is greater or lower than the time windows bounds, it is updated to the nearest bound (Line \ref{SOA:lastCustomerArrivalTime}). Then, the necessary speed $v_\textsc{ref}$ to arrive at time $t'_{\sigma_{e}}$ is computed (Line \ref{SOA:refSpeed}), and the algorithm finds the customer $\sigma_{p}$ with greatest time-window violation when using this speed (Lines \ref{SOA:maxViolation0}-\ref{SOA:maxViolation3}). If no violation is found, the solution is returned. Otherwise, the arrival time $t'_{\sigma_{p}}$ is updated to the nearest time window bound (Line \ref{SOA:updateArrivalTime}) and SOA is called recursively on two subproblems: from $s$ to $p$ (Line \ref{SOA:SOAsubProb1}), and from $p$ to $e$ 
(Line \ref{SOA:SOAsubProb2}).
Once times $t'_{\sigma_{i}}, \forall i = 1, \dots, |\sigma|$, are computed, the associated speeds are revised in such a way that any speed below the optimal speed $v_\textsc{f}^{*}$ which minimizes fuel consumption is replaced by a speed $v_\textsc{f}^{*}$ and a waiting time (Lines \ref{SOA:computeSpeedsBegin}-\ref{SOA:computeSpeedsEnd}). The final arrival dates $t_i$ at route nodes are obtained (Line \ref{SOA:computeSpeedsEnd}).

\begin{algorithm}[htb]
    \caption{Speed Optimization Algorithm --- SOA}
\label{ArrivalTimes}
\begin{algorithmic}[1]
\algsetup{indent=2.5em}
\STATE Procedure $SOA(\sigma,s,e)$ 
\STATE $maxViolation \leftarrow 0$
\STATE $D \leftarrow \sum_{i=s}^{e-1}d_{\sigma_{i},\sigma_{i+1}}$
\STATE $T \leftarrow \sum_{i=s}^{e-1}\tau_{\sigma_{i}}$
\IF {$e = |\sigma|$}
\STATE $t'_{\sigma_{e}} = \min\{\max\{a_{\sigma_{e}},t'_{\sigma_{s}} + D/v^{*}_\textsc{fd} + T\},b_{\sigma_{e}}\}$\label{SOA:lastCustomerArrivalTime}
\ENDIF
\STATE $v_\textsc{ref} \leftarrow D/(t'_{\sigma_{e}} - t'_{\sigma_{s}} - T)$\label{SOA:refSpeed}
\FOR {$i = s+1 \dots e$}
  \STATE $t'_{\sigma_{i}} = t'_{\sigma_{i-1}} + \tau_{\sigma_{i-1}} + d_{\sigma_{i-1},\sigma_{i}}/v_\textsc{ref}$
  \STATE $violation = \max\{0, t'_{\sigma_{i}} - b_{\sigma_{i}}, a_{\sigma_{i}} - t'_{\sigma_{i}}\}$\label{SOA:maxViolation0}
  \IF {$violation > maxViolation$}\label{SOA:maxViolation1}
    \STATE $maxViolation = violation$\label{SOA:maxViolation2}
    \STATE $p = i$\label{SOA:maxViolation3}
  \ENDIF
\ENDFOR
\IF {$maxViolation > 0$}
  \STATE $t'_{\sigma_{p}} = \min\{\max\{a_{\sigma_{p}},t'_{\sigma_{p}} \},b_{\sigma_{p}}\}$\label{SOA:updateArrivalTime}
  \STATE $SOA(\sigma,s,p)$ \label{SOA:SOAsubProb1}
  \STATE $SOA(\sigma,p,e)$\label{SOA:SOAsubProb2}
\ENDIF
\IF {$s = 1$ \textbf{and} $e = |\sigma|$}\label{SOA:computeSpeedsBegin}
\STATE $t_{\sigma_{1}} = 0$
\FOR {$i = 2 \dots |\sigma|$}
  \STATE $v_{\sigma_{i-1},\sigma_{i}} = max\{d_{\sigma_{i-1},\sigma_{i}}/(t'_{\sigma_{i}} - t'_{\sigma_{i-1}} - \tau_{\sigma_{i-1}})), v_\textsc{f}^{*}\}$ \label{SOA:computeSpeeds}
  \STATE$t_{\sigma_{i}} = max\{a_{\sigma_{i-1}}, t_{\sigma_{i-1}}\} + \tau_{\sigma_{i-1}} + d_{\sigma_{i-1},\sigma_{i}}/v_{\sigma_{i-1},\sigma_{i}}$ \label{SOA:computeArrivalTimes}
\ENDFOR
\ENDIF\label{SOA:computeSpeedsEnd}
\end{algorithmic}
\end{algorithm}

\begin{figure}[!ht]
 \begin{center}
  \includegraphics[width=0.76\textwidth]{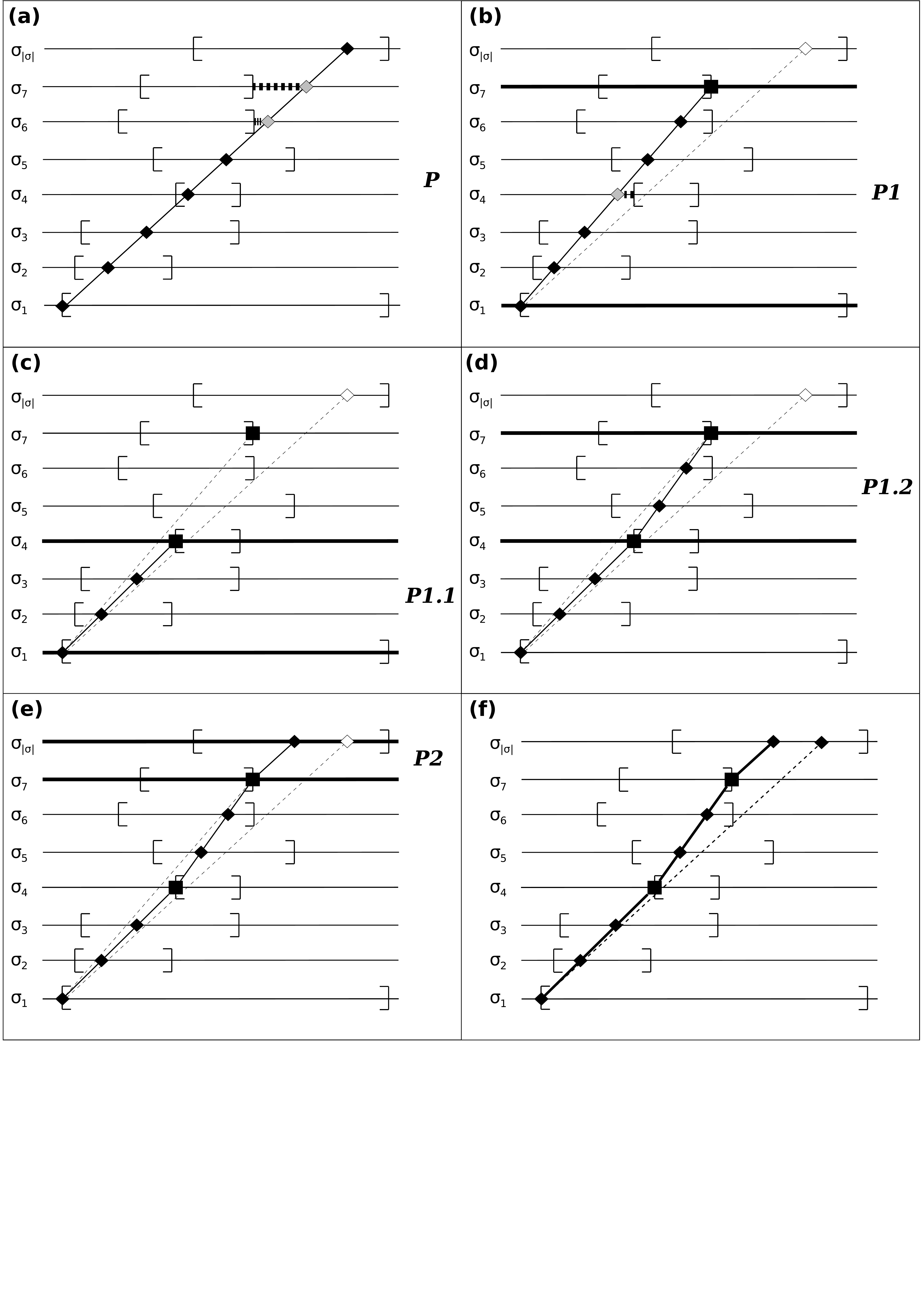}\\
  \caption{Computing arrival times with SOA}
\label{soa_compArrivTimes}
 \end{center}
\end{figure}

Algorithm \ref{ArrivalTimes} is illustrated in Figure \ref{soa_compArrivTimes} on a problem with eight stops. The horizontal axis represents the time, while the vertical axis indicates sequence of customers from bottom to top. The brackets correspond to the time windows and the dots denote the customers' arrival times. Round black dots are related to feasible arrival times; diamond gray dots indicate an early or late arrival time; and square dots indicate the new times after correction. For simplicity, the service times are assumed to be zero.

Figure \ref{soa_compArrivTimes}a illustrates the arrival times at each customer when the vehicle travels with the optimal speed ($v^{*}_\textsc{fd}$) starting from $\sigma_{1} = 0$ at time $t'_{\sigma_{1}} = a_{\sigma_{1}}$. There are two violations and $\sigma_{7}$ is the customer with the greatest violation ($p = 7$).
The arrival time at $\sigma_{7}$ is thus adjusted to $t'_{\sigma_{7}} = b_{\sigma_{7}}$ (Figure \ref{soa_compArrivTimes}b) and two subproblems, P1 and P2, are solved, where P1 considers the subsequence from $\sigma_{1}$ until $\sigma_{7}$, whereas P2 considers the subsequence from $\sigma_{7}$ until $\sigma_{|\sigma|}$.
In P1, the $v_\textsc{ref}$ is computed to arrive at $\sigma_{7}$ at time $t'_{\sigma_{7}}$. A new violation is observed at customer $\sigma_{4}$ ($p = 4$). This violation is corrected (Figure \ref{soa_compArrivTimes}c) and two new subproblems, from $\sigma_{1}$ to $\sigma_{4}$ (P1.1), and from $\sigma_{4}$ to $\sigma_{7}$ (P1.2) are solved. When solving P1.1 and P1.2, no violations are identified  (Figure \ref{soa_compArrivTimes}c-\ref{soa_compArrivTimes}d), and thus the recursion is not applied further. This leads to a feasible solution to P1. Subproblem P2 is solved using almost the same rationale of P1 (see Figure\ref{soa_compArrivTimes}e), except that the customers' arrival times are computed using the optimal speed $v^{*}_\textsc{fd}$ (see Line \ref{SOA:lastCustomerArrivalTime} of Algorithm \ref{ArrivalTimes}). Again, no violation is found, thus completing the whole solution for the problem. At the end, speeds are revised to insert waiting time whenever a speed $v < v^*_\textsc{f}$ is encountered 
(Lines \ref{SOA:computeSpeedsBegin}-\ref{SOA:computeSpeedsEnd}) of Algorithm \ref{ArrivalTimes}).

\section{Computational Results}
\label{Results}

In this section we report the computational experiments for the three variants considered in this work, but first we describe the benchmark instances used for evaluating the performance of our algorithm. 

\subsection{Benchmark Instances}
The first set of instances considered in this work (Set A) is the one from the PRPLIB, suggested by \citet{DemirBL12} and available at \url{http://www.apollo.management.soton.ac.uk/prplib.htm}. This set consists of nine different groups, ranging from 10 to 200 customers, each one containing 20 instances.
For these instances, the same objective parameter values as in \citet{DemirBL12} have been used, that is:

\begin{align*}
 \ w_1 &= 1.01763908\times10^{-3}\\
 \ w_2 &= 5.33605218\times10^{-5}\\
 \ w_3 &= 8.40323178\times10^{-9}\\
 \ w_4 &= 1.41223439\times10^{-7}\\
 \ \omega_{\textsc{fc}} &= 1.4\text{\textsterling}/l\\
 \ \omega_{\textsc{fd}} &= 2.22222222\times10^{-3}\text{\textsterling}/s.
\end{align*}

However, these instances have a large time windows width, such that it is possible to visit many customers within their respective time windows when traveling at optimal speed as further discussed in Section \ref{ResultsPRP}.
In view of this, we created two additional sets of instances with tighter time windows by modifiying those of the PRPLIB. The time horizon of these new sets is $32400$. The time-window width of each customer in Set $B$ is randomly selected between the interval $2000$ and $5000$ with uniform probability, whereas in Set $C$ it is randomly selected between $2000$ and $15000$.
Once the time windows width are defined, the time window lower bound is randomly selected with uniform probability in the time interval $I_i$ which allows to feasibly reach the customer and return to the depot before the end of the time horizon. For a customer $i$, let $W_i$ be the chosen width, and let $\delta_{0i}^\textsc{min} = d_{0i} / v_\textsc{max}$ and $\delta_{i0}^\textsc{min} = d_{i0} / v_\textsc{max}$ be the minimum driving time from and to the depot, respectively. Then, 
\begin{equation}
I_i = [a_0 + \lfloor \delta_{0i_\textsc{min}} \rfloor, b_{0} - \lceil \delta_{i0_\textsc{min}} \rceil - \tau_i - W_i].
\end{equation}

For the FCVRP and the EMVRP, the well-known benchmark instances of \cite{Christofidesetal1979} and \cite{Goldenetal1998} are used. As in \citet{Karaetal2007}, we use $\omega  = 0.15Q$ for the EMVRP.

\subsection{Method performance on the PRP}
\label{ResultsPRP}

The proposed matheuristic algorithm was coded in C++ and executed on an  Intel Core i7 3.40 GHz with 16 GB of RAM, running under Linux Mint 13. 
CPLEX 12.4 was used to solve the SP problems. Only a single thread was used, and the algorithm was run 10 times for each instance with different random seeds.

The following parameter values were adopted: $n_\textsc{r} = 20$, $n_\textsc{ils} = n+5m$, $n_\textsc{sp} = 150$, $n_\textsc{pool} = 2$, $T_\textsc{mip} = 360$ s, and $\omega_\textsc{tw} = 10^{8}$. The first four parameter values directly follow from \cite{Subramanian2012} and \cite{Subramanianetal2013}, whereas the latter is a large number to prevent infeasible solutions due to late arrivals. A higher value of $T_\textsc{mip}$ than \cite{Subramanianetal2013}  is used, as more time seems to be needed by the SP solver to produce improved solutions. 

To investigate the interaction between the local search components, the SP solver, and the speed optimization procedure, we performed further experiments with a static version called ILS-SP-SOA-Stat. For a fair comparison, the CPU time limit for each instance has been set to the average CPU time of the dynamic method for the corresponding instance. In ILS-SP-SOA-Stat, $\mathbf{v}$ does not change during an iteration of the algorithm, i.e., the ``UpdateSpeedMatrix'' function and the \emph{Change Speeds} perturbation is not used in Algorithm (\ref{ILS-SP-SOA}). Hence, the task of intelligently recombining routes with different speed distributions is exclusively operated by the SP solver.

The performance of each method is reported in Table \ref{avgGaps}.  Each line corresponds to averaged results on a set of 20 instances. The columns report for each method and instance set the average Gap(\%) and CPU time. The Gap(\%) for each instance is computed as $100(Z-Z_\textsc{bks})/Z_\textsc{bks}$, where $Z$ is the objective value of the solution and $Z_\textsc{bks}$ is the value of the Best Known Solution (BKS) ever found. Detailed results on each problem instance are provided in \ref{ap:resultsPRP}.

It appears that both versions of the proposed algorithm largely outperform the ALNS heuristic of \citet{DemirBL12} in terms of solution quality. The new methods also appear to be significantly faster than ALNS, with a speed-up ranging from $\times2$ to $\times50$. Note that ALNS experiments have been conducted on a different computing environment and a 3.00GHz CPU. The speed difference between computers of the same generation remains moderate.  It can be observed that instances of Sets B and C require more CPU time for convergence, and solution improvements are found more continuously as the search progresses. This is due to tighter time windows. The increased interplay between route and speed optimization leads to a longer sequence of solution improvements, and thus a slower convergence. This aspect is further developed in the next section.

Comparing the results obtained with the two versions of ILS-SP-SOA, the average solutions obtained with the dynamic version are better than those from the static version in all sets of instances but 200-B and 200-C. The dynamic version examines a wider diversity of speed choices. For smaller instances -- here up to 100 customers -- this allows to converge towards better solutions within the allowed termination criteria. For larger instances with tighter time windows, the benefits of enhanced speed optimization and diversity seems to be counterbalanced by a slower convergence, and the static method, more focused on intensification, may perform marginally better. A longer termination criterion may thus benefit more to the dynamic version that to the static version.

\begin{table}[htbp]
\centering
\caption{Method performance on the PRP benchmark instances}
\small
\scalebox{0.9}
{
\begin{tabular}{cccccccc}
\hline
\textbf{Instance} & \multicolumn{ 3}{c}{\textbf{Average Gap (\%)}} & \textbf{} & \multicolumn{ 3}{c}{\textbf{CPU Time (s)}} \\ \cline{ 2- 4} \cline{ 6- 8}
\textbf{Set} & \textbf{ALNS$^{*}$} & \textbf{ILS-Dynamic} & \textbf{ILS-Static} & \textbf{} & \textbf{ALNS$^{*}$} & \textbf{ILS-Dynamic} & \textbf{ILS-Static} \\ \hline
10-A & 0.03 & \textbf{0.00} & 0.01 &  & 2.34 & 0.04 & 0.04 \\ 
10-B & -- & \textbf{0.01} & 0.11 &  & -- & 0.04 & 0.04 \\ 
10-C & -- & \textbf{0.00} & 0.07 &  & -- & 0.04 & 0.04 \\ \hline
50-A & 0.57 & \textbf{0.01} & 0.09 &  & 35.40 & 2.98 & \ 2.98 \\ 
50-B & -- & \textbf{0.10} & 0.12 &  & -- & 5.26 & 5.26 \\ 
50-C & -- & \textbf{0.23} & 0.32 &  & -- & 4.69 & 4.69 \\ \hline
100-A & 1.99 & \textbf{0.15} & 0.28 &  & 145.27 & 31.98 & 31.98 \\ 
100-B & -- & \textbf{0.16} & 0.26 &  & -- & 91.99 & 91.99 \\ 
100-C & -- & \textbf{0.22} & 0.32 &  & -- & 62.04 & 62.04 \\ \hline
200-A & 4.24 & \textbf{0.54} & 0.59 &  & 625.73 & 296.68 & 296.68 \\ 
200-B & -- & 0.67 & \textbf{0.63} &  & -- & 1162.38 & 1162.38 \\ 
200-C & -- & 0.93 & \textbf{0.92} &  & -- & 533.57 & 533.57 \\ \hline
\multicolumn{7}{l}{\scriptsize{$^{*}$ 3 GHz CPU with 1 GB of RAM}} \vspace*{-0.4cm}
\end{tabular}
\label{avgGaps}
}
\end{table}

%
%

\subsection{Some insights on instance difficulty for the PRP}

To better understand what makes a difficult PRP instance, and how these instances can be better addressed, we further analyze instance characteristics and study how often the optimal speeds are used in the best known solutions. As shown in the following, higher occurrences of ``exotic'' speeds in the BKS, i.e., speeds that are neither $v^{*}_\textsc{f}$ nor $v^{*}_\textsc{fd}$, seem to translate into increased problem difficulty. 

\begin{figure}[htbp]
 \begin{center}
  \includegraphics[width=0.625\textwidth]{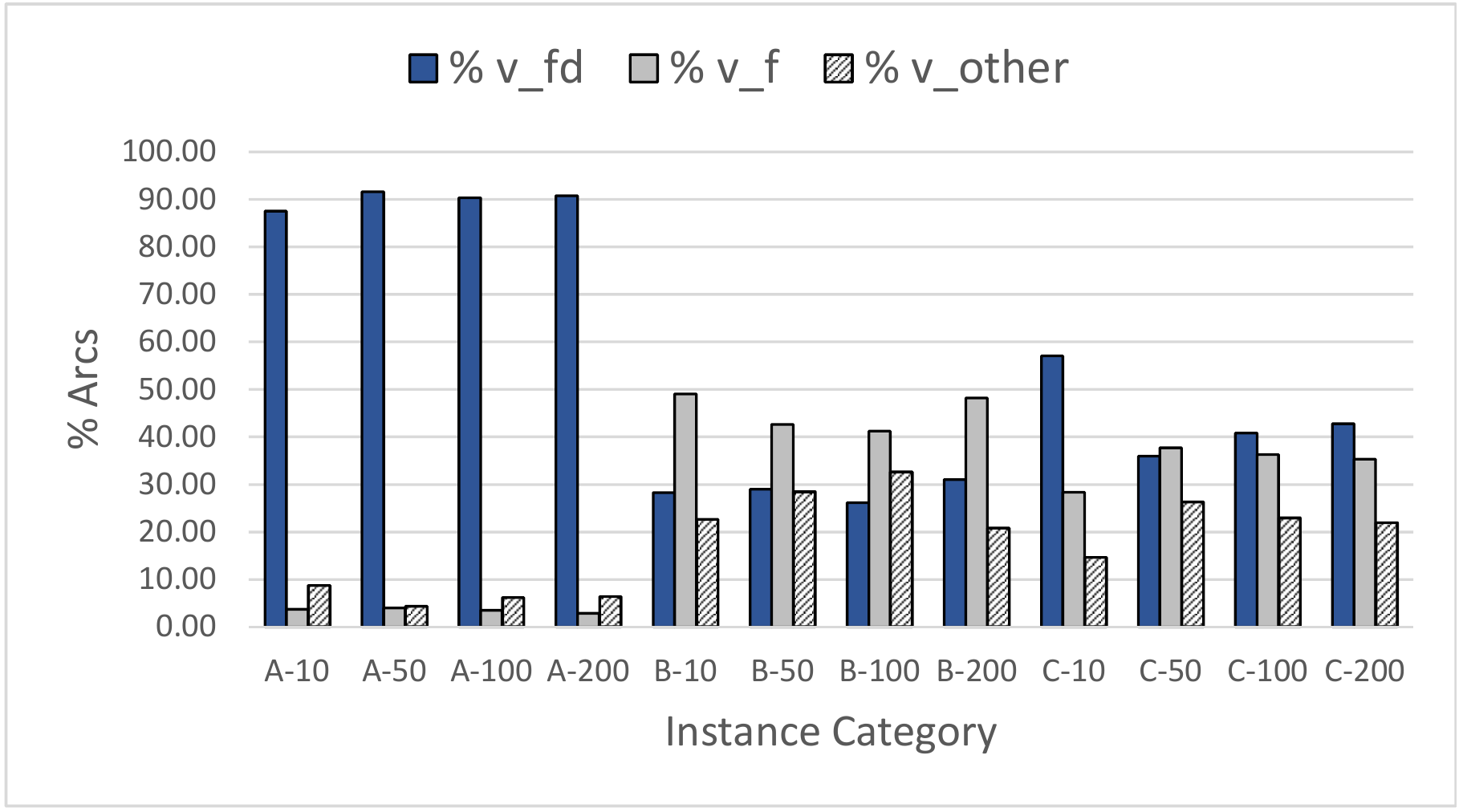}\\
  \caption{Percentage of arcs with  in the BKSs}
\label{ArcsSpeed}
 \end{center}
\end{figure}

Figure \ref{ArcsSpeed} shows the percentage of the number of arcs in which the vehicle travels at a certain speed in the BKS of each group of instances. Three speed categories are discerned: $v^{*}_\textsc{f}$, $v^{*}_\textsc{fd}$, and other speeds $v_\textsc{other}$.
It can be seen that $v^{*}_\textsc{fd}$ appears to be used more often than the other speeds in Set A, regardless of the size of the instance. As expected, $v^{*}_\textsc{f}$ tends to be used much more often in Set B, which is the one with the tightest time windows.  Finally, in Set C, where the time windows are larger than those in Set B,  $v^{*}_\textsc{fd}$ is the most common in the 10-customer instances, 
but there is an equilibrium in the 50-, 100- and 200-customer instances between $v^{*}_\textsc{f}$ and $v^{*}_\textsc{fd}$. 

\begin{figure}[!htbp]
\begin{center}
  \includegraphics[width=0.75\textwidth]{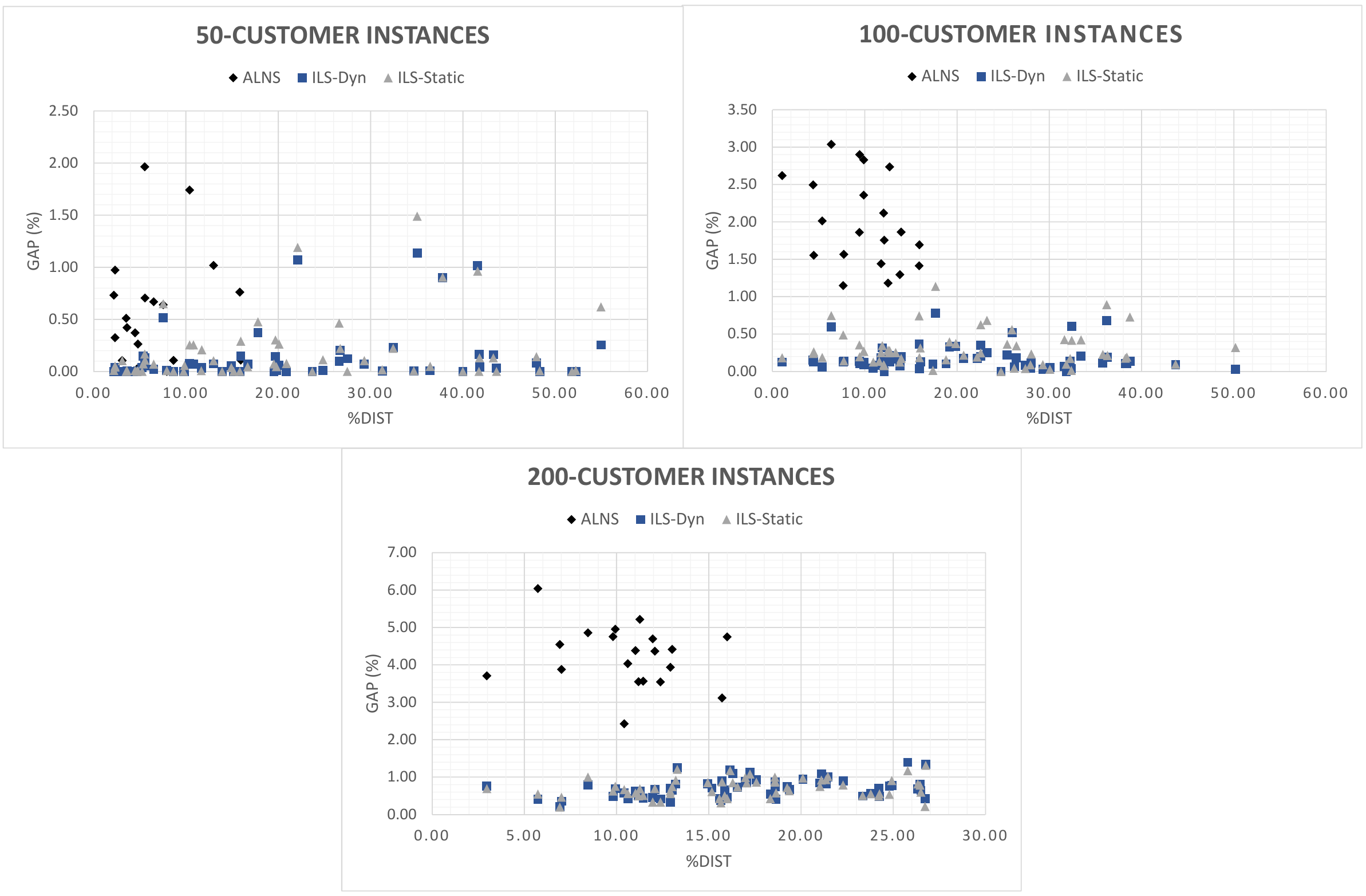}\\
  \caption{Gap (\%) vs Percentage of distance units at other speeds than $v^{*}_\textsc{f}$ and $v^{*}_\textsc{fd}$ in the BKS} \vspace*{-0.3cm}
\label{InstancesGapsDifficulty}
\end{center}
\end{figure}

Let $\%Dist$ be the percentage of the total distance in which the vehicles are traveling with other speeds  than $v^{*}_\textsc{f}$ and $v^{*}_\textsc{fd}$ in the BKS of a given instance. Figure \ref{InstancesGapsDifficulty} displays the average gaps of ILS-SP-SOA-Dyn and ILS-SP-SOA-Stat on the 50-, 100- and 200-customer instances of Sets A, B and C and of ALNS \citep{DemirBL12, DemirThesis2012} on the 50-, 100- and 200-customer instances of Set A, relatively to the $\%Dist$ of the associated instance. 
As noticed earlier, the Gap(\%) obtained by the proposed methods are smaller than those of ALNS. It is also clearly visible that the first set of instances (the only one on which ALNS are reported) required a lower variety of speed values in the BKS. Finally, there is a tendency for larger gaps when $\%Dist$ increases, which may reflect a higher problem difficulty. It should be noted that $v^{*}_\textsc{f}$ and $v_\textsc{other}$ can only arise as a consequence of an active time-window constraint, otherwise $v^{*}_\textsc{fd}$ would be used throughout all the route. Therefore, problems with the largest number of active time-window constraints seem also more difficult.

\begin{figure}[!htbp]
\begin{center}
  \includegraphics[width=0.75\textwidth]{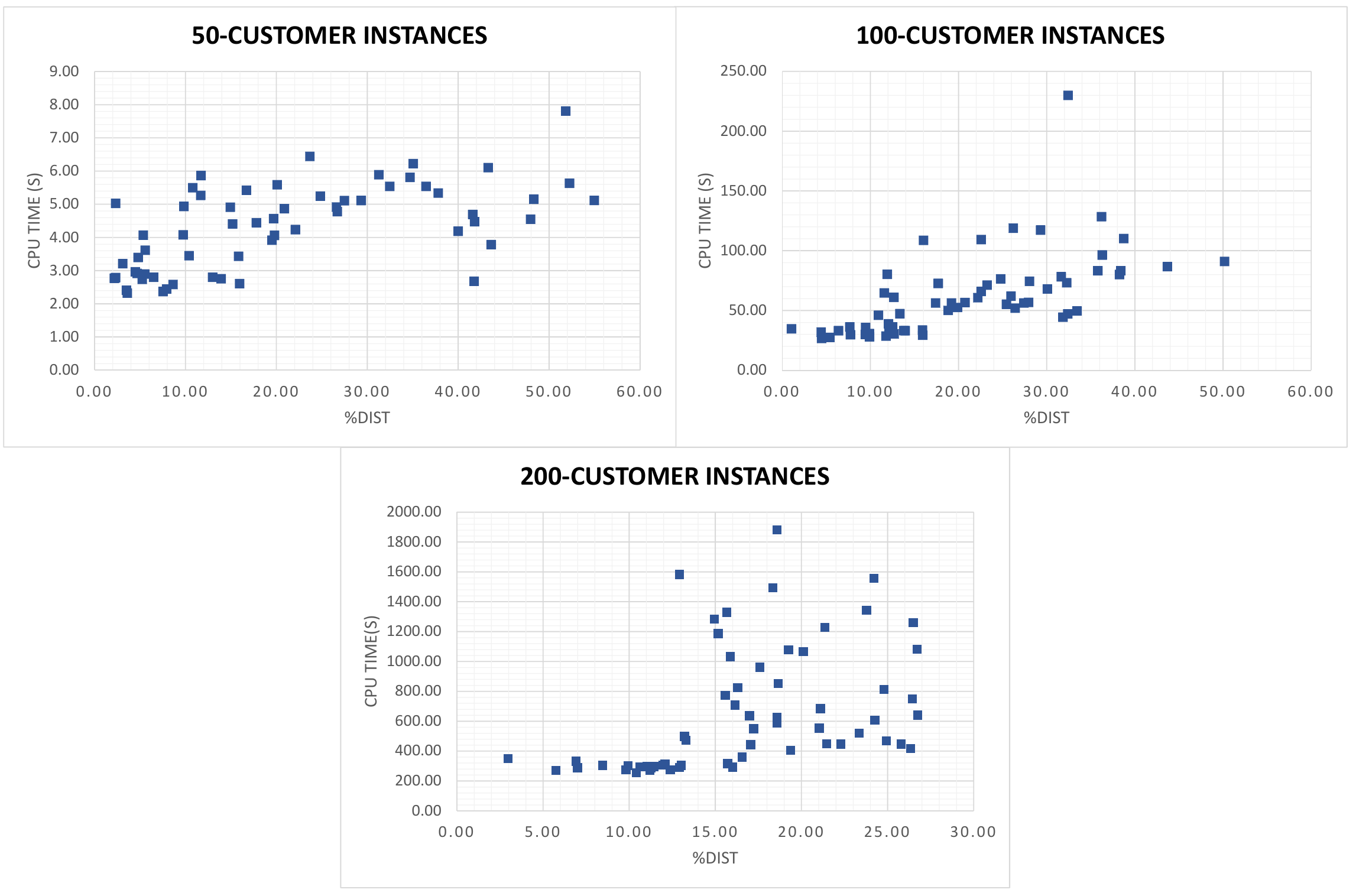}\\
  \caption{Time (s) vs Percentage of distance units at other speeds than $v^{*}_\textsc{f}$ and $v^{*}_\textsc{fd}$ in the BKS} \vspace*{-0.3cm}
\label{InstancesGapsDifficulty2}
\end{center}
\end{figure}

A similar analysis is performed in Figure \ref{InstancesGapsDifficulty2} considering the CPU time of ILS-SP-SOA-Dyn, and thus the convergence speed of the method. Overall, the CPU time needed to converge towards a final solution tends to increase for problems with higher $\%Dist$. This suggests that the instance difficultly is correlated with the need of a variety of speeds in the solutions.

This analysis shows that, even if instances with large time-windows are reasonably well solved by methods with separate phases such as the ALNS of \cite{DemirBL12} or ILS-SP-SOA-Stat, there is still a notable research effort to be done for more intricate problems, when a larger diversity of speeds must be used to fulfill the time constraints. Because of the variety of antagonist interplays between, speed, distance, and feasibility, more integrated methods that consider jointly speed and route optimization within the neighborhoods are likely to be necessary. ILS-SP-SOA-Dyn is already a first step in this direction, and has demonstrated increased performance on some of the most difficult instances.

\subsection{Results on other problems}

%

Computational experiments have been conducted on two additional problems with environmental considerations, the FCVRP and EMVRP, in order to assess the performance of ILS-SP-SOA on a wider variety of settings. 
As previously, a single thread was used and 10 executions were performed for each instance. For a fair comparison with previous methods developed for the FCVRP, we reduced the termination criterion by setting $n_\textsc{r} = 4$, $n_\textsc{ils} = n/5+5m$ and $T_\textsc{mip} = 60$ s. For the EMVRP we only changed the MIP solver time limit to $T_\textsc{mip} = 60$ s. 

\begin{table}[!htb]
\centering
\caption{Results for the FCVRP instances}
\scalebox{0.9}
{
\setlength{\tabcolsep}{1mm}
\begin{tabular}{cc@{\hspace*{0.45cm}}cccc@{\hspace*{0.45cm}}cccccc@{\hspace*{0.45cm}}cc}
\hline
\textbf{} & \textbf{} & \multicolumn{ 3}{c}{\textbf{Avg. Sol. SMSAH}} & \textbf{} & \multicolumn{ 5}{c}{\textbf{Avg. Sol. ILS-SP-SOA}} & \multicolumn{1}{l}{} & \multicolumn{ 2}{c}{\textbf{BKS}} \\ \cline{3-5} \cline{7-11} \cline{13-14}
\textbf{Instance} & \textbf{$n$} & \textbf{} & \textbf{} & \textbf{Gap} & \textbf{} & \textbf{} & \textbf{} & \textbf{} & \textbf{CPU} & \textbf{Gap} & \multicolumn{1}{l}{} & \textbf{} & \textbf{} \\ 
\textbf{} & \textbf{} & \textbf{\up{Cost}} & \textbf{\up{Dist.}} & \textbf{(\%)} & \textbf{} & \textbf{\up{Cost}} & \textbf{\up{Dist.}} & \textbf{\up{$|\mathbf{R}|$}} & \textbf{time (s)} & \textbf{(\%)} & \multicolumn{1}{l}{} & \textbf{\up{Cost}} & \textbf{\up{$|\mathbf{R}|$}} \\ \hline
C1 & 50 & 751.43 & 534.14 & 0.04 &  & 757.86 & 536.45 & 5.00 & 0.21 & 0.90 &  & \textbf{751.11} & 5 \\ 
C2 & 75 & 1188.62 & 862.60 & 1.36 &  & 1182.28 & 864.86 & 10.00 & 1.04 & 0.82 &  & \underline{\textbf{1172.62}} & 10 \\ 
C3 & 100 & 1153.56 & 849.45 & 0.50 &  & 1151.00 & 852.61 & 8.00 & 1.81 & 0.28 &  & \textbf{1147.83} & 8 \\ 
C4 & 150 & 1461.69 & 1066.11 & 1.04 &  & 1450.85 & 1064.72 & 12.00 & 5.39 & 0.29 &  & \underline{\textbf{1446.64}} & 12 \\ 
C5 & 199 & 1865.30 & 1348.88 & 1.55 &  & 1850.03 & 1349.46 & 17.00 & 13.30 & 0.72 &  & \underline{\textbf{1836.86}} & 17 \\ 
C6 & 50 & -- & -- & -- &  & 1255.54 & 565.92 & 6.00 & 0.31 & 0.01 &  & \textbf{1255.39} & 6 \\ 
C7 & 75 & -- & -- & -- &  & 1988.58 & 927.08 & 11.00 & 1.67 & 0.08 &  & \textbf{1986.91} & 11 \\ 
C8 & 100 & -- & -- & -- &  & 2168.89 & 881.43 & 9.00 & 2.95 & 0.04 &  & \textbf{2168.01} & 9 \\ 
C9 & 150 & -- & -- & -- &  & 3052.07 & 1184.41 & 14.00 & 16.01 & 0.14 &  & \textbf{3047.77} & 14 \\ 
C10 & 199 & -- & -- & -- &  & 3909.95 & 1437.95 & 18.00 & 43.59 & 0.18 &  & \textbf{3902.99} & 18 \\ 
C11 & 120 & 1516.42 & 1054.06 & 0.19 &  & 1543.64 & 1080.23 & 7.00 & 4.49 & 1.99 &  & 1513.48$^{1}$ & -- \\ 
C12 & 100 & 1175.59 & 827.98 & 0.13 &  & 1174.02 & 827.05 & 10.00 & 1.12 & 0.00 &  & \textbf{1174.02} & 10 \\ 
C13 & 120 & -- & -- & -- &  & 7984.38 & 1567.07 & 11.00 & 33.26 & 0.14 &  & \textbf{7973.52} & 11 \\ 
C14 & 100 & -- & -- & -- &  & 10213.40 & 868.68 & 11.00 & 1.71 & 0.00 &  & \textbf{10213.40} & 11 \\ \hline
\textbf{Avg} &  &  &  & \textbf{2.21} &  &  &  &  & \textbf{9.06} & \textbf{0.40} &  &  &  \\ \hline
G1 & 240 & 7714.29 & 5706.19 & 0.70 &  & 7663.90 & 5862.75 & 10.00 & 50.21 & 0.04 &  & \underline{\textbf{7660.64}} & 10 \\ 
G2 & 320 & 11195.02 & 8490.24 & 0.39 &  & 11166.00 & 8501.47 & 10.00 & 200.13 & 0.13 &  & \underline{\textbf{11151.20}} & 10 \\ 
G3 & 400 & 14566.73 & 11097.24 & 0.58 &  & 14485.57 & 11092.66 & 10.00 & 365.81 & 0.02 &  & \underline{\textbf{14482.60}} & 10 \\ 
G4 & 480 & 18605.37 & 13848.72 & 2.18 &  & 18265.40 & 14132.20 & 12.00 & 389.43 & 0.31 &  & \underline{\textbf{18209.30}} & 12 \\ 
G5 & 200 & 8576.91 & 6471.57 & 0.18 &  & 8561.53 & 6460.98 & 5.00 & 21.89 & 0.00 &  & \textbf{8561.53} & 5 \\ 
G6 & 280 & 11121.04 & 8431.39 & 0.42 &  & 11089.13 & 8599.16 & 8.00 & 68.62 & 0.13 &  & \underline{\textbf{11074.40}} & 8 \\ 
G7 & 360 & 13477.07 & 10249.70 & 0.63 &  & 13393.74 & 10242.70 & 9.00 & 238.36 & 0.01 &  & \underline{\textbf{13392.90}} & 9 \\ 
G8 & 440 & 16098.60 & 11888.02 & 3.92 &  & 15491.72 & 11875.90 & 11.00 & 599.50 & 0.00 &  & \underline{\textbf{15491.30}} & 11 \\ 
G9 & 255 & 858.34 & 612.79 & 2.47 &  & 845.08 & 609.82 & 14.00 & 36.97 & 0.89 &  & \underline{\textbf{837.63}} & 14 \\ 
G10 & 323 & 1090.85 & 777.10 & 2.43 &  & 1069.92 & 772.39 & 16.00 & 73.32 & 0.47 &  & \underline{\textbf{1064.97}} & 16 \\ 
G11 & 399 & 1360.20 & 971.48 & 2.87 &  & 1328.85 & 961.56 & 18.00 & 185.27 & 0.49 &  & \underline{\textbf{1322.31}} & 18 \\ 
G12 & 483 & 1661.07 & 1182.80 & 3.73 &  & 1612.76 & 1158.42 & 19.00 & 337.01 & 0.71 &  & \underline{\textbf{1601.35}} & 19 \\ 
G13 & 252 & 1269.37 & 904.67 & 2.70 &  & 1241.11 & 895.13 & 26.90 & 20.09 & 0.41 &  & \underline{\textbf{1236.03}} & 27 \\ 
G14 & 320 & 1604.83 & 1141.03 & 2.52 &  & 1573.20 & 1131.46 & 30.00 & 38.52 & 0.50 &  & \underline{\textbf{1565.44}} & 30 \\ 
G15 & 396 & 1987.76 & 1417.55 & 2.65 &  & 1943.26 & 1400.23 & 34.00 & 70.82 & 0.35 &  & \underline{\textbf{1936.45}} & 34 \\ 
G16 & 480 & 2408.72 & 1714.48 & 2.89 &  & 2352.11 & 1697.73 & 37.10 & 152.74 & 0.47 &  & \underline{\textbf{2341.08}} & 37 \\ 
G17 & 240 & 1033.88 & 725.33 & 1.56 &  & 1025.50 & 722.48 & 22.00 & 22.17 & 0.73 &  & \underline{\textbf{1018.02}} & 22 \\ 
G18 & 300 & 1469.97 & 1032.93 & 1.67 &  & 1454.61 & 1033.07 & 28.00 & 39.05 & 0.60 &  & \underline{\textbf{1445.88}} & 28 \\ 
G19 & 360 & 2014.26 & 1420.35 & 2.09 &  & 1987.13 & 1413.10 & 33.00 & 106.19 & 0.72 &  & \underline{\textbf{1972.97}} & 33 \\ 
G20 & 420 & 2699.29 & 1903.12 & 2.46 &  & 2660.40 & 1898.12 & 39.00 & 122.89 & 0.99 &  & \underline{\textbf{2634.42}} & 39 \\ \hline
\textbf{Avg} &  &  &  & \textbf{1.95} &  &  &  &  & \textbf{156.95} & \textbf{0.40} &  &  &  \\ \hline
\multicolumn{14}{l}{\small{$^{1}$Result from \cite{Xiaoetal2012}}}\\
\end{tabular}
\label{resultsFCVRP}
}
\end{table}

For the FCVRP, the results obtained by ILS-SP-SOA were compared with those of \cite{Xiaoetal2012}, as can be observed in Table \ref{resultsFCVRP}. The values highlighted only in boldface indicate that the BKS found by our algorithm equaled the best one reported by \cite{Xiaoetal2012}, whereas those in boldface and underline indicate that the BKS found by ILS-SP-SOA improved the one presented by the same authors. 
To our knowledge, no results on the instances of \cite{Christofidesetal1979} were available for the EMVRP. Hence, Table \ref{resultsEMVRP} reports the gap between the average solutions and the BKS found by ILS-SP-SOA.

\begin{table}[!htb]
\centering
\renewcommand{\arraystretch}{0.92}
\caption{Results for the EMVRP instances}
\scalebox{0.9}
{
\setlength{\tabcolsep}{2mm}
\begin{tabular}{cc@{\hspace*{0.45cm}}ccccc@{\hspace*{0.45cm}}ccccc@{\hspace*{0.45cm}}}
\hline
\multicolumn{ 1}{c}{} & \multicolumn{ 1}{c}{} & \multicolumn{ 5}{c}{\textbf{Avg. Sol. ILS-SP-SOA (10 runs)}} & \textbf{} & \multicolumn{ 4}{c}{\textbf{Best Sol. ILS-SP-SOA (10 runs)}} \\ \cline{ 3- 7} \cline{ 9- 12}
\multicolumn{ 1}{c}{\textbf{Instance}} & \multicolumn{ 1}{c}{\textbf{$n$}} & \textbf{} & \textbf{} & \textbf{} & \textbf{CPU} & \textbf{Gap} & \textbf{} & \textbf{} & \textbf{} & \textbf{} & \textbf{CPU} \\
\multicolumn{ 1}{c}{} & \multicolumn{ 1}{c}{} & \textbf{\up{Cost}} & \textbf{\up{Dist.}} & \textbf{\up{$|\mathbf{R}|$}} & \textbf{T(s)} & \textbf{(\%)} & \textbf{} & \textbf{\up{Cost}} & \textbf{\up{Dist.}} & \textbf{\up{$|\mathbf{R}|$}} & \textbf{T(s)} \\ \hline
C1 & 50 & 46383.05 & 564.04 & 5.00 & 0.91 & 0.37 &  & 46210.35 & 564.75 & 5 & 0.97 \\ 
C2 & 75 & 60575.76 & 885.88 & 10.00 & 3.71 & 0.02 &  & 60564.99 & 883.89 & 10 & 3.68 \\ 
C3 & 100 & 83490.41 & 883.17 & 8.00 & 8.03 & 0.00 &  & 83490.41 & 883.17 & 8 & 8.76 \\ 
C4 & 10 & 105869.55 & 1121.89 & 12.00 & 22.88 & 0.03 &  & 105842.20 & 1123.03 & 12 & 23.49 \\ 
C5 & 199 & 135142.92 & 1441.77 & 17.00 & 48.14 & 0.01 &  & 135123.96 & 1454.19 & 17 & 42.99 \\ 
C6 & 50 & 42403.25 & 604.90 & 6.00 & 1.33 & 0.00 &  & 42403.25 & 604.90 & 6 & 1.54 \\ 
C7 & 75 & 61164.47 & 965.81 & 11.00 & 5.67 & 0.00 &  & 61164.47 & 965.81 & 11 & 5.89 \\ 
C8 & 100 & 80247.16 & 944.25 & 9.00 & 14.00 & 0.00 &  & 80247.16 & 944.25 & 9 & 14.32 \\ 
C9 & 150 & 105618.73 & 1235.21 & 14.00 & 95.29 & 0.32 &  & 105281.84 & 1235.05 & 14 & 71.35 \\ 
C10 & 199 & 136390.40 & 1513.39 & 18.00 & 119.93 & 0.42 &  & 135819.48 & 1524.32 & 18 & 132.36 \\ 
C11 & 120 & 123430.83 & 1098.82 & 7.00 & 27.81 & 0.07 &  & 123348.14 & 1100.81 & 7 & 26.45 \\ 
C12 & 100 & 93521.30 & 859.54 & 10.00 & 5.80 & 0.00 &  & 93521.30 & 859.54 & 10 & 5.06 \\ 
C13 & 120 & 128356.05 & 1603.45 & 11.00 & 76.65 & 0.01 &  & 128345.69 & 1602.06 & 11 & 111.48 \\ 
C14 & 100 & 93827.15 & 922.69 & 11.00 & 10.68 & 0.00 &  & 93827.15 & 922.69 & 11 & 10.67 \\ \hline
\textbf{Avg.} &  &  &  &  & \textbf{31.49} & \textbf{0.09} &  &  &  &  &  \\ \hline
G1 & 240 & 1451439.74 & 6000.85 & 10.00 & 281.01 & 0.00 &  & 1451439.72 & 6000.85 & 10 & 305.62 \\ 
G2 & 320 & 2736274.33 & 8599.93 & 10.00 & 1288.77 & 0.01 &  & 2736086.81 & 8601.13 & 10 & 1228.78 \\ 
G3 & 400 & 4527654.48 & 11189.40 & 10.00 & 2007.98 & 0.00 &  & 4527654.42 & 11189.40 & 10 & 2101.69 \\ 
G4 & 480 & 6150983.46 & 14459.29 & 12.00 & 2395.43 & 0.08 &  & 6146280.48 & 14452.60 & 12 & 2430.96 \\ 
G5 & 200 & 2742484.44 & 6536.42 & 5.00 & 100.62 & 0.00 &  & 2742484.44 & 6536.42 & 5 & 102.71 \\ 
G6 & 280 & 3352707.87 & 8793.30 & 8.00 & 380.48 & 0.01 &  & 3352282.80 & 8794.36 & 8 & 404.98 \\ 
G7 & 360 & 4205444.72 & 10336.80 & 9.00 & 1305.71 & 0.00 &  & 4205444.71 & 10336.80 & 9 & 1363.80 \\ 
G8 & 440 & 4830980.29 & 11971.71 & 11.00 & 2397.67 & 0.04 &  & 4828825.54 & 11970.00 & 11 & 2819.51 \\ 
G9 & 255 & 302320.91 & 707.76 & 14.00 & 162.71 & 0.06 &  & 302127.64 & 705.51 & 14 & 146.63 \\ 
G10 & 323 & 384217.14 & 908.24 & 16.00 & 325.81 & 0.04 &  & 384069.09 & 917.59 & 16 & 347.21 \\ 
G11 & 399 & 476933.10 & 1133.93 & 18.00 & 646.64 & 0.04 &  & 476743.98 & 1128.05 & 18 & 644.06 \\ 
G12 & 483 & 579327.34 & 1327.30 & 19.00 & 1401.90 & 0.48 &  & 576582.81 & 1326.01 & 19 & 930.49 \\ 
G13 & 252 & 448193.11 & 1007.59 & 27.00 & 89.00 & 0.07 &  & 447899.37 & 1004.09 & 27 & 81.55 \\ 
G14 & 320 & 567382.89 & 1273.73 & 30.00 & 186.63 & 0.10 &  & 566803.22 & 1277.20 & 30 & 135.04 \\ 
G15 & 396 & 700357.75 & 1596.06 & 34.00 & 394.35 & 0.11 &  & 699609.07 & 1596.33 & 34 & 460.52 \\ 
G16 & 480 & 846679.46 & 1952.54 & 38.00 & 675.41 & 0.08 &  & 845985.34 & 1947.01 & 38 & 584.14 \\ 
G17 & 240 & 80393.13 & 772.07 & 22.00 & 82.65 & 0.09 &  & 80319.35 & 768.26 & 22 & 77.46 \\ 
G18 & 300 & 111393.13 & 1132.17 & 28.00 & 182.85 & 0.07 &  & 111317.98 & 1131.81 & 28 & 175.26 \\ 
G19 & 360 & 151155.20 & 1559.64 & 33.00 & 319.42 & 0.21 &  & 150845.06 & 1557.55 & 33 & 312.78 \\ 
G20 & 420 & 199052.22 & 2160.16 & 40.60 & 502.77 & 0.04 &  & 198976.65 & 2176.22 & 41 & 459.86 \\ \hline
\textbf{Avg.} &  &  &  &  & \textbf{756.39} & \textbf{0.08} &  &  &  &  &  \\ \hline
\end{tabular}
\label{resultsEMVRP}
}
\end{table}

These results show that ILS-SP-SOA consistently produces solutions of high quality on these two problems. We observe that 22 new BKS solutions were found for the FCVRP, and the average gaps between the average solutions of ILS-SP-SOA and the BKSs is $0.16\%$ for the set of instances of \cite{Christofidesetal1979} and $0.31\%$ for the one of \cite{Goldenetal1998}. This gap is much smaller than \cite{Xiaoetal2012}, which attained gaps of $2.20\%$ and $1.95\%$.
Note that the seven instances that include service times and route duration constraints, namely C6-C10 and C13-C14 were not solved by \cite{Xiaoetal2012}. When comparing the solution cost between each of the seven pair of instances, i.e, C1 and C6; C2 and C7; C3 and C8; C4 and C9; C5 and C10; C11 and C13; C12 and C14; where the first instance of the pair does not consider service times and route duration constraints, while the second one does, it can be observed that the solution costs increase dramatically when these characteristics are considered.
Regarding the CPU time, \cite{Xiaoetal2012} only reported the mean of the average times, which was approximately 78 s for the set of instances of \cite{Christofidesetal1979} and 198 s for the one of \cite{Goldenetal1998}. Our algorithm spent, on average, 9.06 s and 156.95 s, respectively, as shown in Table \ref{resultsFCVRP}. It should be pointed out that \cite{Xiaoetal2012} ran their experiments on a slower 1.6 GHz notebook.

For the EMVRP, and in contrast to the FCVRP, service times and route duration constraints do not have a dramatic impact in the solution cost when compared to the associated instances that do not consider these characteristics.  
Overall, the gaps between the average solutions and the BKS are small, more precisely $0.09\%$ and $0.08\%$, on the instances of \cite{Christofidesetal1979} and \cite{Goldenetal1998}, respectively. When observing such a small variance for VRP problems we can reasonably conjecture that the problem is fairly-well solved.

\section{Concluding remarks}
\label{Conclusions}

This paper dealt with the Pollution-Routing Problem (PRP), a ``green''-oriented variant of the Vehicle Routing Problem (VRP). In order to solve it, we proposed a matheuristic approach, called ILS-SOA-SP, that effectively integrates Iterated Local Search (ILS) with a Set Partitioning (SP) procedure and a Speed Optimization Algorithm (SOA). This approach was also used to solve two other environmental based VRPs, namely the Fuel Consumption Vehicle Routing Problem (FCVRP) and the Energy Minimizing Vehicle Routing Problem (EMVRP). The results of extensive computational experiments demonstrate that ILS-SP-SOA is capable of generating high-quality solutions in a very consistent manner, outperforming previous methods from the literature.

Two new sets of PRP instances were introduced, and our computational experiments show that they are more challenging than the previous one available in the literature. We also showed that the level of difficulty of an instance tends to increase when the vehicles are likely to travel at speeds other than the optimal ones. In other words, good (or even feasible) routing solutions are not easily obtained without using non-optimal speeds, and choosing the correct ones can be a challenging task. The proposed method has the advantage to  perform joint route and speed optimization within several local search and integer programming components, and thus performs much better on difficult instances than previously available algorithms.

As a promising line of research, we believe that further efforts are still necessary to incorporate speed decisions more tightly within the local search. This can be done, for example, by developing larger neighborhood structures that are capable of exploring promising regions of the solution space by considering different speeds during the local search. The exploration of these combined neighborhoods is likely to be highly time-consuming, and thus very clever move selection and move evaluation methods may need to be developed.

\bibliography{references}
\newpage

\appendix

\section{Fuel consumption model}
\label{sec::parameters}

The fuel consumption model used in the PRP formulation, given by Eq. (\ref{fuelConsumptionModel}), is based on the comprehensive emissions model described by \citet{Barthetal2005}, \citet{ScoraBarth2006} and \citet{BarthBoriboonsomsin2008}. It is given by
\begin{align}
 \ Z_{ij}(v_{ij}) = \lambda ( kNV + w \gamma \alpha_{ij} v_{ij} + \gamma \alpha_{ij} f_{ij} v_{ij} + \beta \gamma v_{ij}^3 ) d_{ij} / v_{ij}, \label{fuelConsumptionModel}
\end{align}
where $\lambda = \xi / \kappa \psi$ and $\gamma = 1/1000 \eta_{tf} \eta$ are constants related to the fuel properties, $\beta = 0,5 C_d \rho A$ and $w$ are constants associated with the vehicle characteristics and $\alpha_{ij}$ is constant that depends on the road characteristics and on the vehicle acceleration. More precisely, $\alpha_{ij} = \tau_{ij} + g \sin \theta_{ij} + g C_r \cos \theta_{ij}$, where $\tau_{ij}$ is the acceleration and $\theta_{ij}$ is the road angle inclination. The meaning and value of all parameters is given in Table \ref{prp:parameters}.

Assuming that acceleration and road inclination are null, the fuel consumption can alternatively be written as 
\begin{equation}
Z_{ij}(v_{ij}) = (w_1/v_{ij} + w_2 + w_3 f_{ij} + w_4v_{ij}^{2})d_{ij},
\end{equation}
where $w_1 = \lambda kNV$, $w_2 = \lambda w \gamma g C_r$,  $w_3 = \lambda \gamma g C_r$ and $w_4 = \lambda \beta \gamma$. \\

\vspace{-0.25cm}
\begin{table}[htbp!]
\caption{Parameters used to solve the PRP}
\centering
\small
\scalebox{0.9}
{
\begin{tabular}{lll}
\hline
Notation & Description \\ \hline
$w$ & Curb-weight ($kg$) & $6350$ \\
$\xi$ & Fuel-to-air mass ratio & $1$ \\
$k$ & Engine friction factor ($kJ/rev/l$) & $0.2$ \\
$N$ & Engine speed ($rev/s$) & $33$ \\
$V$ & Engine displacement ($l$) & $5$ \\
$g$ & Gravitacional constant ($m/s^2$) & $9.81$ \\
$C_d$ & Coefficient of aerodynamic drag & $0.7$ \\
$\rho$ & Air density ($kg/m^3$) & $1.2041$ \\
$A$ & Frontal surface area ($m^2$) & $3.912$ \\
$C_r$ & Coefficient of rolling resistence & $0.01$ \\
$\eta_{tf}$ & Vehicle drive train efficiency & $0.4$ \\
$\eta$ & Efficiency parameter for diesel engines & $0.9$ \\
$\omega_{\textsc{fc}}$ & Fuel and CO$_2$ emissions cost per liter (\textsterling)  & $1.4$ \\
$\omega_{\textsc{fd}}$ & Driver wage (\textsterling$/s$) & $8/3600$ \\
$\kappa$ & Heating value of a typical diesel fuel ($kJ/g$) & $44$ \\
$\psi$ & Conversion factor ($g/s$ para $l/s$) & $737$ \\
$v_\textsc{min}$ & Lower speed limit ($m/s$) & $5.5$ (or $20 km/h$) \\
$v_\textsc{max}$ & Upper speed limit ($m/s$) & $25$ (or $90 km/h$) \\ \hline
\multicolumn{ 3}{l}{Source: \citep{DemirBL12}}
\end{tabular}
\label{prp:parameters}
}
\end{table}
\vspace{-0.4cm}
\section{Results for the PRP}
\label{ap:resultsPRP}

In the tables presented hereafter, \emph{Instance} denotes the name of the instance, \emph{Cost} indicates the value of the objective function, $|\mathbf{R}|$ is the number of vehicles, \emph{CPU T(s)} represents the computing time in seconds, \emph{Gap(\%)} is the gap between the solution cost and CPLEX for the 10-customer instances or the best known solution obtained among all experiments for the other instances, given in the column \emph{BKS}. Note that the approximate optimal solutions for the 10-customer instances were computed with the MIP formulation of \citet{DemirBL12}. It is impractical to consider all possible speed values in the model, and thus a speed-discretization was operated with 500 levels ranging from $v^{*}_\textsc{f}$ to $v^{*}_\textsc{fd}$. This leads in general to very high-quality upper bounds.

\begin{table}[htbp]
\centering
\renewcommand{\arraystretch}{0.92}
\caption{Results for the PRP 10-customer instances}
\small
\scalebox{0.9}
{
\setlength{\tabcolsep}{1mm}
\begin{tabular}{ccccccc@{\hspace*{0.4cm}}ccccc@{\hspace*{0.4cm}}ccccc@{\hspace*{0.4cm}}cc}
\hline
\textbf{} & \textbf{} & \multicolumn{ 4}{c}{\textbf{ALNS}} &  & \multicolumn{4}{c}{\textbf{Avg. ILS-SP-SOA-Dyn}} &  & \multicolumn{ 4}{c}{\textbf{Avg. ILS-SP-SOA-Stat}} &  & \multicolumn{ 2}{c}{\textbf{CPLEX}} \\ \cline{ 3- 6} \cline{ 8- 11}\cline{ 13- 16}\cline{ 18- 19}
\textbf{Instance} & \textbf{} & \textbf{} & \textbf{} & \textbf{CPU} & \textbf{Gap} & \textbf{} & \textbf{} & \textbf{} & \textbf{CPU} & \textbf{Gap} & \textbf{} & \textbf{} & \textbf{} & \textbf{CPU} & \textbf{Gap} & \textbf{} & \textbf{} & \textbf{} \\
\multicolumn{ 1}{c}{} & \textbf{} & \textbf{\up{Cost}} & \textbf{\up{$|\mathbf{R}|$}} & \textbf{T(s)$^{*}$} & \textbf{(\%)} & \textbf{} & \textbf{\up{Cost}} & \textbf{\up{$|\mathbf{R}|$}} & \textbf{T(s)} & \textbf{(\%)} & \textbf{} & \textbf{\up{Cost}} & \textbf{\up{$|\mathbf{R}|$}} & \textbf{T(s)} & \textbf{(\%)} & \textbf{} & \textbf{\up{Cost}} & \textbf{\up{$|\mathbf{R}|$}} \\ \hline
UK10\_01 &  & 170.64 & 2 & 2.1 & 0.00 &  & 170.64 & 2.0 & 0.04 & 0.00 &  & 170.64 & 2.0 & 0.04 & 0.00 &  & \textbf{170.64} & 2 \\ 
UK10\_02 &  & 204.88 & 2 & 2.3 & 0.00 &  & 204.88 & 2.0 & 0.05 & 0.00 &  & 204.88 & 2.0 & 0.05 & 0.00 &  & \textbf{204.88} & 2 \\ 
UK10\_03 &  & 200.42 & 3 & 2.0 & 0.04 &  & 200.40 & 3.0 & 0.03 & 0.03 &  & 200.62 & 3.0 & 0.03 & 0.14 &  & \textbf{200.34} & 3 \\ 
UK10\_04 &  & 189.99 & 2 & 2.2 & 0.05 &  & 189.88 & 2.0 & 0.04 & 0.00 &  & 189.88 & 2.0 & 0.04 & 0.00 &  & \textbf{189.89} & 2 \\ 
UK10\_05 &  & 175.59 & 2 & 2.3 & 0.00 &  & 175.59 & 2.0 & 0.04 & 0.00 &  & 175.59 & 2.0 & 0.04 & 0.00 &  & \textbf{175.59} & 2 \\ 
UK10\_06 &  & 214.48 & 2 & 2.2 & 0.00 &  & 214.48 & 2.0 & 0.05 & 0.00 &  & 214.48 & 2.0 & 0.05 & 0.00 &  & \textbf{214.48} & 2 \\ 
UK10\_07 &  & 190.14 & 2 & 2.9 & 0.00 &  & 190.14 & 2.0 & 0.04 & 0.00 &  & 190.14 & 2.0 & 0.04 & 0.00 &  & \textbf{190.14} & 2 \\ 
UK10\_08 &  & 222.17 & 2 & 2.1 & 0.00 &  & 222.17 & 2.0 & 0.03 & 0.00 &  & 222.17 & 2.0 & 0.03 & 0.00 &  & \textbf{222.17} & 2 \\ 
UK10\_09 &  & 174.54 & 2 & 2.2 & 0.00 &  & 174.54 & 2.0 & 0.04 & 0.00 &  & 174.54 & 2.0 & 0.04 & 0.00 &  & \textbf{174.54} & 2 \\ 
UK10\_10 &  & 190.04 & 2 & 2.6 & 0.12 &  & 189.82 & 2.0 & 0.04 & 0.00 &  & 189.82 & 2.0 & 0.04 & 0.00 &  & \textbf{189.82} & 2 \\ 
UK10\_11 &  & 262.08 & 2 & 2.2 & 0.00 &  & 262.08 & 2.0 & 0.03 & 0.00 &  & 262.08 & 2.0 & 0.03 & 0.00 &  & \textbf{262.08} & 2 \\ 
UK10\_12 &  & 183.19 & 2 & 2.2 & 0.00 &  & 183.19 & 2.0 & 0.04 & 0.00 &  & 183.19 & 2.0 & 0.04 & 0.00 &  & \textbf{183.19} & 2 \\ 
UK10\_13 &  & 195.97 & 2 & 2.2 & 0.00 &  & 195.97 & 2.0 & 0.04 & 0.00 &  & 195.97 & 2.0 & 0.04 & 0.00 &  & \textbf{195.97} & 2 \\ 
UK10\_14 &  & 163.28 & 2 & 2.4 & 0.07 &  & 163.17 & 2.0 & 0.04 & 0.00 &  & 163.17 & 2.0 & 0.04 & 0.00 &  & \textbf{163.17} & 2 \\ 
UK10\_15 &  & 127.24 & 2 & 2.4 & 0.11 &  & 127.10 & 2.0 & 0.05 & 0.00 &  & 127.10 & 2.0 & 0.05 & 0.00 &  & \textbf{127.10} & 2 \\ 
UK10\_16 &  & 186.73 & 2 & 1.9 & 0.05 &  & 186.63 & 2.0 & 0.04 & 0.00 &  & 186.63 & 2.0 & 0.04 & 0.00 &  & \textbf{186.63} & 2 \\ 
UK10\_17 &  & 159.03 & 2 & 2.3 & 0.00 &  & 159.03 & 2.0 & 0.04 & 0.00 &  & 159.03 & 2.0 & 0.04 & 0.00 &  & \textbf{159.03} & 2 \\ 
UK10\_18 &  & 162.09 & 2 & 2.2 & 0.00 &  & 162.09 & 2.0 & 0.04 & 0.00 &  & 162.09 & 2.0 & 0.04 & 0.00 &  & \textbf{162.09} & 2 \\ 
UK10\_19 &  & 169.59 & 2 & 4.1 & 0.07 &  & 169.46 & 2.0 & 0.04 & 0.00 &  & 169.46 & 2.0 & 0.04 & 0.00 &  & \textbf{169.46} & 2 \\ 
UK10\_20 &  & 168.80 & 2 & 2.0 & 0.00 &  & 168.80 & 2.0 & 0.03 & 0.00 &  & 168.80 & 2.0 & 0.03 & 0.00 &  & \textbf{168.80} & 2 \\ \hline
\textbf{Avg.} &  &  &  & \textbf{2.34} & \textbf{0.03} &  &  &  & \textbf{0.04} & \textbf{0.00} &  &  &  & \textbf{0.04} & \textbf{0.01} &  &  &  \\ \hline
UK10\_01-B &  & -- & -- & -- & -- &  & 246.45 & 2.0 & 0.05 & 0.00 &  & 246.45 & 2.0 & 0.05 & 0.00 &  & \textbf{246.45} & 2 \\ 
UK10\_02-B &  & -- & -- & -- & -- &  & 303.73 & 2.0 & 0.05 & 0.00 &  & 303.73 & 2.0 & 0.05 & 0.00 &  & \textbf{303.73} & 2 \\ 
UK10\_03-B &  & -- & -- & -- & -- &  & 301.89 & 3.0 & 0.04 & 0.00 &  & 301.89 & 3.0 & 0.04 & 0.00 &  & \textbf{301.89} & 3 \\ 
UK10\_04-B &  & -- & -- & -- & -- &  & 273.90 & 2.0 & 0.04 & 0.00 &  & 273.90 & 2.0 & 0.04 & 0.00 &  & \textbf{273.91} & 2 \\ 
UK10\_05-B &  & -- & -- & -- & -- &  & 255.07 & 2.0 & 0.06 & 0.00 &  & 255.07 & 2.0 & 0.06 & 0.00 &  & \textbf{255.08} & 2 \\ 
UK10\_06-B &  & -- & -- & -- & -- &  & 332.34 & 3.0 & 0.04 & 0.00 &  & 332.34 & 3.0 & 0.04 & 0.00 &  & \textbf{332.34} & 3 \\ 
UK10\_07-B &  & -- & -- & -- & -- &  & 314.64 & 3.0 & 0.04 & 0.00 &  & 314.64 & 3.0 & 0.04 & 0.00 &  & \textbf{314.64} & 3 \\ 
UK10\_08-B &  & -- & -- & -- & -- &  & 339.36 & 2.0 & 0.06 & 0.00 &  & 339.36 & 2.0 & 0.06 & 0.00 &  & \textbf{339.37} & 2 \\ 
UK10\_09-B &  & -- & -- & -- & -- &  & 261.10 & 2.0 & 0.04 & 0.00 &  & 261.10 & 2.0 & 0.04 & 0.00 &  & \textbf{261.10} & 2 \\ 
UK10\_10-B &  & -- & -- & -- & -- &  & 285.20 & 2.0 & 0.04 & 0.00 &  & 285.20 & 2.0 & 0.04 & 0.00 &  & \textbf{285.20} & 2 \\ 
UK10\_11-B &  & -- & -- & -- & -- &  & 409.39 & 3.0 & 0.04 & 0.20 &  & 410.66 & 3.0 & 0.04 & 0.52 &  & \textbf{408.55} & 3 \\ 
UK10\_12-B &  & -- & -- & -- & -- &  & 251.65 & 2.0 & 0.05 & 0.00 &  & 252.18 & 2.0 & 0.05 & 0.21 &  & \textbf{251.65} & 2 \\ 
UK10\_13-B &  & -- & -- & -- & -- &  & 274.07 & 3.0 & 0.05 & 0.00 &  & 274.07 & 3.0 & 0.05 & 0.00 &  & \textbf{274.08} & 3 \\ 
UK10\_14-B &  & -- & -- & -- & -- &  & 267.92 & 2.0 & 0.04 & 0.00 &  & 267.92 & 2.0 & 0.04 & 0.00 &  & \textbf{267.92} & 2 \\ 
UK10\_15-B &  & -- & -- & -- & -- &  & 197.50 & 2.0 & 0.06 & 0.00 &  & 197.50 & 2.0 & 0.06 & 0.00 &  & \textbf{197.50} & 2 \\ 
UK10\_16-B &  & -- & -- & -- & -- &  & 245.76 & 2.0 & 0.03 & 0.00 &  & 245.76 & 2.0 & 0.03 & 0.00 &  & \textbf{245.76} & 2 \\ 
UK10\_17-B &  & -- & -- & -- & -- &  & 283.83 & 2.0 & 0.04 & 0.00 &  & 283.83 & 2.0 & 0.04 & 0.00 &  & \textbf{283.83} & 2 \\ 
UK10\_18-B &  & -- & -- & -- & -- &  & 241.53 & 2.0 & 0.05 & 0.00 &  & 245.00 & 2.0 & 0.05 & 1.44 &  & \textbf{241.53} & 2 \\ 
UK10\_19-B &  & -- & -- & -- & -- &  & 330.38 & 3.0 & 0.04 & 0.00 &  & 330.38 & 3.0 & 0.04 & 0.00 &  & \textbf{330.38} & 3 \\ 
UK10\_20-B &  & -- & -- & -- & -- &  & 208.06 & 2.0 & 0.04 & 0.00 &  & 208.06 & 2.0 & 0.04 & 0.00 &  & \textbf{208.06} & 2 \\ \hline
\textbf{Avg.} &  &  &  & \textbf{--} & \textbf{--} &  &  &  & \textbf{0.04} & \textbf{0.01} &  &  &  & \textbf{0.04} & \textbf{0.11} &  &  &  \\ \hline
UK10\_01-C &  & -- & -- & -- & -- &  & 210.18 & 2.0 & 0.04 & 0.00 &  & 210.18 & 2.0 & 0.04 & 0.00 &  & \textbf{210.18} & 2 \\ 
UK10\_02-C &  & -- & -- & -- & -- &  & 271.93 & 2.0 & 0.05 & 0.00 &  & 271.93 & 2.0 & 0.05 & 0.00 &  & \textbf{271.93} & 2 \\ 
UK10\_03-C &  & -- & -- & -- & -- &  & 229.18 & 2.0 & 0.04 & 0.00 &  & 229.18 & 2.0 & 0.04 & 0.00 &  & \textbf{229.18} & 2 \\ 
UK10\_04-C &  & -- & -- & -- & -- &  & 230.59 & 2.0 & 0.04 & 0.00 &  & 230.59 & 2.0 & 0.04 & 0.00 &  & \textbf{230.59} & 2 \\ 
UK10\_05-C &  & -- & -- & -- & -- &  & 205.49 & 2.0 & 0.04 & 0.00 &  & 205.49 & 2.0 & 0.04 & 0.00 &  & \textbf{205.49} & 2 \\ 
UK10\_06-C &  & -- & -- & -- & -- &  & 255.82 & 2.0 & 0.04 & 0.00 &  & 255.82 & 2.0 & 0.04 & 0.00 &  & \textbf{255.82} & 2 \\ 
UK10\_07-C &  & -- & -- & -- & -- &  & 217.79 & 2.0 & 0.03 & 0.00 &  & 217.79 & 2.0 & 0.03 & 0.00 &  & \textbf{217.79} & 2 \\ 
UK10\_08-C &  & -- & -- & -- & -- &  & 251.29 & 2.0 & 0.04 & 0.00 &  & 251.29 & 2.0 & 0.04 & 0.00 &  & \textbf{251.29} & 2 \\ 
UK10\_09-C &  & -- & -- & -- & -- &  & 186.04 & 2.0 & 0.03 & 0.00 &  & 186.04 & 2.0 & 0.03 & 0.00 &  & \textbf{186.04} & 2 \\ 
UK10\_10-C &  & -- & -- & -- & -- &  & 231.62 & 2.0 & 0.05 & 0.00 &  & 231.62 & 2.0 & 0.05 & 0.00 &  & \textbf{231.62} & 2 \\ 
UK10\_11-C &  & -- & -- & -- & -- &  & 298.20 & 2.0 & 0.04 & 0.00 &  & 298.20 & 2.0 & 0.04 & 0.00 &  & \textbf{298.20} & 2 \\ 
UK10\_12-C &  & -- & -- & -- & -- &  & 206.58 & 2.0 & 0.04 & 0.00 &  & 208.54 & 2.0 & 0.04 & 0.95 &  & \textbf{206.58} & 2 \\ 
UK10\_13-C &  & -- & -- & -- & -- &  & 211.75 & 2.0 & 0.04 & 0.00 &  & 211.75 & 2.0 & 0.04 & 0.00 &  & \textbf{211.75} & 2 \\ 
UK10\_14-C &  & -- & -- & -- & -- &  & 209.07 & 2.0 & 0.04 & 0.00 &  & 209.07 & 2.0 & 0.04 & 0.00 &  & \textbf{209.07} & 2 \\ 
UK10\_15-C &  & -- & -- & -- & -- &  & 176.56 & 2.0 & 0.05 & 0.00 &  & 176.56 & 2.0 & 0.05 & 0.00 &  & \textbf{176.56} & 2 \\ 
UK10\_16-C &  & -- & -- & -- & -- &  & 229.15 & 2.0 & 0.04 & 0.00 &  & 229.15 & 2.0 & 0.04 & 0.00 &  & \textbf{229.15} & 2 \\ 
UK10\_17-C &  & -- & -- & -- & -- &  & 219.20 & 2.0 & 0.04 & 0.00 &  & 219.20 & 2.0 & 0.04 & 0.00 &  & \textbf{219.20} & 2 \\ 
UK10\_18-C &  & -- & -- & -- & -- &  & 195.04 & 2.0 & 0.05 & 0.00 &  & 195.90 & 2.0 & 0.05 & 0.44 &  & \textbf{195.04} & 2 \\ 
UK10\_19-C &  & -- & -- & -- & -- &  & 218.19 & 2.0 & 0.05 & 0.00 &  & 218.19 & 2.0 & 0.05 & 0.00 &  & \textbf{218.19} & 2 \\ 
UK10\_20-C &  & -- & -- & -- & -- &  & 189.56 & 2.0 & 0.04 & 0.00 &  & 189.56 & 2.0 & 0.04 & 0.00 &  & \textbf{189.56} & 2 \\ \hline
\textbf{Avg.} &  &  &  & \textbf{--} & \textbf{--} &  &  &  & \textbf{0.04} & \textbf{0.00} &  &  &  & \textbf{0.04} & \textbf{0.07} &  &  &  \\ \hline
\multicolumn{10}{l}{\scriptsize{$^{*}$ 3 GHz CPU with 1 GB of RAM}}
\end{tabular}
\label{resultsPRP10}
}
\end{table}
\begin{table}[htbp]
\centering
\renewcommand{\arraystretch}{0.92}
\caption{Results for the PRP 50-customer instances}
\small
\scalebox{0.9}
{
\setlength{\tabcolsep}{1mm}
\begin{tabular}{ccccccc@{\hspace*{0.4cm}}ccccc@{\hspace*{0.4cm}}ccccc@{\hspace*{0.4cm}}cc}
\hline
\textbf{} & \textbf{} & \multicolumn{ 4}{c}{\textbf{ALNS}} &  & \multicolumn{4}{c}{\textbf{Avg. ILS-SP-SOA-Dyn}} &  & \multicolumn{ 4}{c}{\textbf{Avg. ILS-SP-SOA-Stat}} &  & \multicolumn{ 2}{c}{\textbf{BKS}} \\ \cline{ 3- 6} \cline{ 8- 11}\cline{ 13- 16}\cline{ 18- 19}
\textbf{Instance} & \textbf{} & \textbf{} & \textbf{} & \textbf{CPU} & \textbf{Gap} & \textbf{} & \textbf{} & \textbf{} & \textbf{CPU} & \textbf{Gap} & \textbf{} & \textbf{} & \textbf{} & \textbf{CPU} & \textbf{Gap} & \textbf{} & \textbf{} & \textbf{} \\
\multicolumn{ 1}{c}{} & \textbf{} & \textbf{\up{Cost}} & \textbf{\up{$|\mathbf{R}|$}} & \textbf{T(s)$^{*}$} & \textbf{(\%)} & \textbf{} & \textbf{\up{Cost}} & \textbf{\up{$|\mathbf{R}|$}} & \textbf{T(s)} & \textbf{(\%)} & \textbf{} & \textbf{\up{Cost}} & \textbf{\up{$|\mathbf{R}|$}} & \textbf{T(s)} & \textbf{(\%)} & \textbf{} & \textbf{\up{Cost}} & \textbf{\up{$|\mathbf{R}|$}} \\ \hline
UK50\_01 &  & 593.77 & 7 & 29.70 & 0.11 &  & 593.14 & 7.0 & 2.58 & 0.00 &  & 593.14 & 7.0 & 2.58 & 0.00 &  & \textbf{593.14} & 7 \\ 
UK50\_02 &  & 599.43 & 7 & 60.30 & 0.11 &  & 599.66 & 7.0 & 2.61 & 0.15 &  & 600.52 & 7.0 & 2.61 & 0.29 &  & \textbf{598.78} & 7 \\ 
UK50\_03 &  & 626.21 & 7 & 53.40 & 0.73 &  & 621.66 & 7.0 & 2.77 & 0.00 &  & 621.72 & 7.0 & 2.77 & 0.01 &  & \textbf{621.66} & 7 \\ 
UK50\_04 &  & 740.92 & 8 & 35.80 & 0.37 &  & 738.18 & 8.0 & 2.97 & 0.00 &  & 738.25 & 8.0 & 2.97 & 0.01 &  & \textbf{738.18} & 8 \\ 
UK50\_05 &  & 636.00 & 6 & 35.30 & 0.51 &  & 632.77 & 6.0 & 2.41 & 0.00 &  & 632.77 & 6.0 & 2.41 & 0.00 &  & \textbf{632.77} & 6 \\ 
UK50\_06 &  & 584.61 & 8 & 54.60 & 0.02 &  & 584.47 & 8.0 & 2.92 & 0.00 &  & 584.47 & 8.0 & 2.92 & 0.00 &  & \textbf{584.47} & 8 \\ 
UK50\_07 &  & 541.07 & 7 & 25.70 & 0.76 &  & 536.98 & 7.0 & 3.43 & 0.00 &  & 536.98 & 7.0 & 3.43 & 0.00 &  & \textbf{536.98} & 7 \\ 
UK50\_08 &  & 560.27 & 7 & 39.50 & 0.32 &  & 558.66 & 7.0 & 2.79 & 0.04 &  & 558.75 & 7.0 & 2.79 & 0.05 &  & \textbf{558.46} & 7 \\ 
UK50\_09 &  & 687.79 & 7 & 21.40 & 0.67 &  & 683.38 & 7.0 & 2.80 & 0.02 &  & 683.70 & 7.0 & 2.80 & 0.07 &  & \textbf{683.22} & 7 \\ 
UK50\_10 &  & 670.92 & 7 & 25.60 & 0.97 &  & 664.58 & 7.0 & 5.03 & 0.02 &  & 664.73 & 7.0 & 5.03 & 0.04 &  & \textbf{664.45} & 7 \\ 
UK50\_11 &  & 618.94 & 7 & 25.10 & 0.11 &  & 618.29 & 7.0 & 3.21 & 0.00 &  & 618.93 & 7.0 & 3.21 & 0.10 &  & \textbf{618.29} & 7 \\ 
UK50\_12 &  & 571.42 & 7 & 40.70 & 0.64 &  & 570.72 & 6.4 & 2.37 & 0.52 &  & 571.47 & 6.0 & 2.37 & 0.65 &  & \textbf{567.78} & 6 \\ 
UK50\_13 &  & 589.11 & 7 & 52.10 & 0.42 &  & 586.68 & 7.0 & 2.32 & 0.01 &  & 586.69 & 7.0 & 2.32 & 0.01 &  & \textbf{586.64} & 7 \\ 
UK50\_14 &  & 660.17 & 7 & 35.00 & 0.71 &  & 655.90 & 7.0 & 3.62 & 0.05 &  & 656.25 & 7.0 & 3.62 & 0.11 &  & \textbf{655.54} & 7 \\ 
UK50\_15 &  & 584.13 & 6 & 26.20 & 0.02 &  & 584.24 & 6.0 & 2.74 & 0.04 &  & 584.02 & 6.0 & 2.74 & 0.00 &  & \textbf{584.02} & 6 \\ 
UK50\_16 &  & 585.16 & 7 & 51.00 & 1.96 &  & 574.63 & 7.0 & 2.90 & 0.13 &  & 574.85 & 7.0 & 2.90 & 0.17 &  & \textbf{573.89} & 7 \\ 
UK50\_17 &  & 456.56 & 7 & 20.00 & 0.00 &  & 456.61 & 7.0 & 2.45 & 0.01 &  & 456.60 & 7.0 & 2.45 & 0.01 &  & \textbf{456.56} & 7 \\ 
UK50\_18 &  & 681.72 & 8 & 26.40 & 0.26 &  & 679.95 & 8.0 & 3.40 & 0.00 &  & 679.96 & 8.0 & 3.40 & 0.00 &  & \textbf{679.93} & 8 \\ 
UK50\_19 &  & 597.95 & 7 & 21.20 & 1.74 &  & 588.17 & 7.0 & 3.45 & 0.08 &  & 589.22 & 7.0 & 3.45 & 0.26 &  & \textbf{587.72} & 7 \\ 
UK50\_20 &  & 678.56 & 7 & 28.90 & 1.02 &  & 672.21 & 7.0 & 2.80 & 0.07 &  & 672.41 & 7.0 & 2.80 & 0.10 &  & \textbf{671.72} & 7 \\ \hline
\textbf{Avg.} & \textbf{} & \textbf{} & \textbf{} & \textbf{35.40} & \textbf{0.57} & \textbf{} & \textbf{} & \textbf{} & \textbf{2.98} & \textbf{0.06} & \textbf{} & \textbf{} & \textbf{} & \textbf{2.98} & \textbf{0.09} & \textbf{} & \textbf{} & \textbf{} \\ \hline
UK50\_01-B &  & -- & -- & -- & -- &  & 882.73 & 7.0 & 4.94 & 0.00 &  & 883.24 & 7.0 & 4.94 & 0.06 &  & \textbf{882.73} & 7 \\ 
UK50\_02-B &  & -- & -- & -- & -- &  & 866.88 & 7.0 & 5.11 & 0.12 &  & 865.82 & 7.0 & 5.11 & 0.00 &  & \textbf{865.82} & 7 \\ 
UK50\_03-B &  & -- & -- & -- & -- &  & 857.74 & 7.0 & 4.91 & 0.06 &  & 857.59 & 7.0 & 4.91 & 0.04 &  & \textbf{857.26} & 7 \\ 
UK50\_04-B &  & -- & -- & -- & -- &  & 993.27 & 8.0 & 5.90 & 0.00 &  & 993.41 & 8.0 & 5.90 & 0.02 &  & \textbf{993.25} & 8 \\ 
UK50\_05-B &  & -- & -- & -- & -- &  & 877.56 & 7.0 & 5.16 & 0.00 &  & 877.61 & 7.0 & 5.16 & 0.01 &  & \textbf{877.56} & 7 \\ 
UK50\_06-B &  & -- & -- & -- & -- &  & 831.33 & 8.0 & 5.54 & 0.23 &  & 831.25 & 8.0 & 5.54 & 0.22 &  & \textbf{829.41} & 8 \\ 
UK50\_07-B &  & -- & -- & -- & -- &  & 747.87 & 7.0 & 4.41 & 0.00 &  & 747.87 & 7.0 & 4.41 & 0.00 &  & \textbf{747.87} & 7 \\ 
UK50\_08-B &  & -- & -- & -- & -- &  & 806.21 & 7.0 & 7.81 & 0.00 &  & 806.21 & 7.0 & 7.81 & 0.00 &  & \textbf{806.21} & 7 \\ 
UK50\_09-B &  & -- & -- & -- & -- &  & 923.82 & 7.0 & 5.87 & 0.04 &  & 925.40 & 7.0 & 5.87 & 0.21 &  & \textbf{923.48} & 7 \\ 
UK50\_10-B &  & -- & -- & -- & -- &  & 895.81 & 7.4 & 5.34 & 0.90 &  & 895.86 & 7.5 & 5.34 & 0.90 &  & \textbf{887.83} & 7 \\ 
UK50\_11-B &  & -- & -- & -- & -- &  & 878.13 & 7.0 & 6.11 & 0.16 &  & 877.86 & 7.0 & 6.11 & 0.13 &  & \textbf{876.71} & 7 \\ 
UK50\_12-B &  & -- & -- & -- & -- &  & 780.27 & 6.5 & 5.12 & 0.25 &  & 783.12 & 6.7 & 5.12 & 0.62 &  & \textbf{778.29} & 6 \\ 
UK50\_13-B &  & -- & -- & -- & -- &  & 802.43 & 7.0 & 3.92 & 0.00 &  & 803.01 & 7.0 & 3.92 & 0.07 &  & \textbf{802.43} & 7 \\ 
UK50\_14-B &  & -- & -- & -- & -- &  & 943.68 & 7.0 & 5.27 & 0.03 &  & 943.46 & 7.0 & 5.27 & 0.01 &  & \textbf{943.39} & 7 \\ 
UK50\_15-B &  & -- & -- & -- & -- &  & 814.68 & 7.0 & 3.78 & 0.03 &  & 814.41 & 7.0 & 3.78 & 0.00 &  & \textbf{814.41} & 7 \\ 
UK50\_16-B &  & -- & -- & -- & -- &  & 781.63 & 7.0 & 6.44 & 0.00 &  & 781.63 & 7.0 & 6.44 & 0.00 &  & \textbf{781.63} & 7 \\ 
UK50\_17-B &  & -- & -- & -- & -- &  & 707.17 & 7.0 & 4.07 & 0.15 &  & 706.65 & 7.0 & 4.07 & 0.07 &  & \textbf{706.14} & 7 \\ 
UK50\_18-B &  & -- & -- & -- & -- &  & 907.73 & 8.0 & 4.08 & 0.00 &  & 907.73 & 8.0 & 4.08 & 0.00 &  & \textbf{907.73} & 8 \\ 
UK50\_19-B &  & -- & -- & -- & -- &  & 807.68 & 7.0 & 5.81 & 0.00 &  & 807.71 & 7.0 & 5.81 & 0.01 &  & \textbf{807.66} & 7 \\ 
UK50\_20-B &  & -- & -- & -- & -- &  & 909.44 & 7.0 & 5.63 & 0.00 &  & 909.54 & 7.0 & 5.63 & 0.01 &  & \textbf{909.44} & 7 \\ \hline
\textbf{Avg.} & \textbf{} & \textbf{} & \textbf{} & \textbf{--} & \textbf{--} & \textbf{} & \textbf{} & \textbf{} & \textbf{5.26} & \textbf{0.10} & \textbf{} & \textbf{} & \textbf{} & \textbf{5.26} & \textbf{0.12} & \textbf{} & \textbf{} & \textbf{} \\ \hline
UK50\_01-C &  & -- & -- & -- & -- &  & 775.44 & 7.0 & 5.25 & 0.01 &  & 776.23 & 7.0 & 5.25 & 0.11 &  & \textbf{775.36} & 7 \\ 
UK50\_02-C &  & -- & -- & -- & -- &  & 755.42 & 7.0 & 4.48 & 0.04 &  & 756.10 & 7.0 & 4.48 & 0.13 &  & \textbf{755.11} & 7 \\ 
UK50\_03-C &  & -- & -- & -- & -- &  & 765.09 & 8.0 & 4.87 & 0.00 &  & 765.68 & 8.0 & 4.87 & 0.08 &  & \textbf{765.09} & 8 \\ 
UK50\_04-C &  & -- & -- & -- & -- &  & 871.90 & 7.3 & 6.23 & 1.14 &  & 874.94 & 7.3 & 6.23 & 1.49 &  & \textbf{862.10} & 7 \\ 
UK50\_05-C &  & -- & -- & -- & -- &  & 789.42 & 6.0 & 4.55 & 0.08 &  & 789.89 & 6.0 & 4.55 & 0.14 &  & \textbf{788.77} & 6 \\ 
UK50\_06-C &  & -- & -- & -- & -- &  & 739.75 & 8.0 & 5.11 & 0.07 &  & 740.00 & 8.0 & 5.11 & 0.10 &  & \textbf{739.23} & 8 \\ 
UK50\_07-C &  & -- & -- & -- & -- &  & 703.96 & 7.0 & 4.70 & 1.01 &  & 703.60 & 6.9 & 4.70 & 0.96 &  & \textbf{696.89} & 6 \\ 
UK50\_08-C &  & -- & -- & -- & -- &  & 686.74 & 7.0 & 5.54 & 0.01 &  & 687.03 & 7.0 & 5.54 & 0.05 &  & \textbf{686.69} & 7 \\ 
UK50\_09-C &  & -- & -- & -- & -- &  & 819.75 & 7.0 & 4.19 & 0.00 &  & 819.75 & 7.0 & 4.19 & 0.00 &  & \textbf{819.75} & 7 \\ 
UK50\_10-C &  & -- & -- & -- & -- &  & 820.65 & 7.0 & 5.43 & 0.07 &  & 820.46 & 7.0 & 5.43 & 0.05 &  & \textbf{820.08} & 7 \\ 
UK50\_11-C &  & -- & -- & -- & -- &  & 739.76 & 7.0 & 5.59 & 0.06 &  & 741.27 & 7.0 & 5.59 & 0.26 &  & \textbf{739.32} & 7 \\ 
UK50\_12-C &  & -- & -- & -- & -- &  & 686.68 & 6.0 & 2.68 & 0.16 &  & 685.55 & 6.0 & 2.68 & 0.00 &  & \textbf{685.55} & 6 \\ 
UK50\_13-C &  & -- & -- & -- & -- &  & 759.84 & 7.0 & 4.92 & 0.10 &  & 762.60 & 7.0 & 4.92 & 0.46 &  & \textbf{759.08} & 7 \\ 
UK50\_14-C &  & -- & -- & -- & -- &  & 788.20 & 7.0 & 4.06 & 0.01 &  & 788.48 & 7.0 & 4.06 & 0.05 &  & \textbf{788.12} & 7 \\ 
UK50\_15-C &  & -- & -- & -- & -- &  & 717.98 & 6.0 & 2.76 & 0.00 &  & 717.98 & 6.0 & 2.76 & 0.00 &  & \textbf{717.98} & 6 \\ 
UK50\_16-C &  & -- & -- & -- & -- &  & 724.44 & 6.1 & 4.24 & 1.07 &  & 725.30 & 6.0 & 4.24 & 1.19 &  & \textbf{716.76} & 6 \\ 
UK50\_17-C &  & -- & -- & -- & -- &  & 625.51 & 7.0 & 4.45 & 0.37 &  & 626.16 & 7.0 & 4.45 & 0.48 &  & \textbf{623.19} & 7 \\ 
UK50\_18-C &  & -- & -- & -- & -- &  & 833.14 & 8.0 & 4.57 & 0.14 &  & 834.45 & 8.0 & 4.57 & 0.30 &  & \textbf{831.94} & 8 \\ 
UK50\_19-C &  & -- & -- & -- & -- &  & 727.20 & 7.0 & 4.78 & 0.20 &  & 727.34 & 7.0 & 4.78 & 0.22 &  & \textbf{725.74} & 7 \\ 
UK50\_20-C &  & -- & -- & -- & -- &  & 812.55 & 7.0 & 5.50 & 0.07 &  & 814.04 & 7.0 & 5.50 & 0.25 &  & \textbf{811.97} & 7 \\ \hline
\textbf{Avg.} & \textbf{} & \textbf{} & \textbf{} & \textbf{--} & \textbf{--} & \textbf{} & \textbf{} & \textbf{} & \textbf{4.69} & \textbf{0.23} & \textbf{} & \textbf{} & \textbf{} & \textbf{4.69} & \textbf{0.32} & \textbf{} & \textbf{} & \textbf{} \\ \hline
\multicolumn{10}{l}{\scriptsize{$^{*}$ 3 GHz CPU with 1 GB of RAM}}
\end{tabular}
\label{resultsPRP50}
}
\end{table}
\begin{table}[htbp]
\centering
\renewcommand{\arraystretch}{0.92}
\caption{Results for the PRP 100-customer instances}
\small
\scalebox{0.9}
{
\setlength{\tabcolsep}{1mm}
\begin{tabular}{ccccccc@{\hspace*{0.4cm}}ccccc@{\hspace*{0.4cm}}ccccc@{\hspace*{0.4cm}}cc}
\hline
\textbf{} & \textbf{} & \multicolumn{ 4}{c}{\textbf{ALNS}} &  & \multicolumn{4}{c}{\textbf{Avg. ILS-SP-SOA-Dyn}} &  & \multicolumn{ 4}{c}{\textbf{Avg. ILS-SP-SOA-Stat}} &  & \multicolumn{ 2}{c}{\textbf{BKS}} \\ \cline{ 3- 6} \cline{ 8- 11}\cline{ 13- 16}\cline{ 18- 19}
\textbf{Instance} & \textbf{} & \textbf{} & \textbf{} & \textbf{CPU} & \textbf{Gap} & \textbf{} & \textbf{} & \textbf{} & \textbf{CPU} & \textbf{Gap} & \textbf{} & \textbf{} & \textbf{} & \textbf{CPU} & \textbf{Gap} & \textbf{} & \textbf{} & \textbf{} \\
\multicolumn{ 1}{c}{} & \textbf{} & \textbf{\up{Cost}} & \textbf{\up{$|\mathbf{R}|$}} & \textbf{T(s)$^{*}$} & \textbf{(\%)} & \textbf{} & \textbf{\up{Cost}} & \textbf{\up{$|\mathbf{R}|$}} & \textbf{T(s)} & \textbf{(\%)} & \textbf{} & \textbf{\up{Cost}} & \textbf{\up{$|\mathbf{R}|$}} & \textbf{T(s)} & \textbf{(\%)} & \textbf{} & \textbf{\up{Cost}} & \textbf{\up{$|\mathbf{R}|$}} \\ \hline
UK100\_01 &  & 1240.79 & 14 & 92.10 & 2.62 &  & 1210.63 & 14.0 & 34.65 & 0.13 &  & 1211.34 & 14.0 & 34.65 & 0.18 &  & \textbf{1209.11} & 14 \\ 
UK100\_02 &  & 1168.17 & 13 & 98.20 & 1.86 &  & 1149.07 & 13.0 & 33.20 & 0.20 &  & 1148.79 & 13.0 & 33.20 & 0.17 &  & \textbf{1146.79} & 13 \\ 
UK100\_03 &  & 1092.73 & 13 & 207.90 & 1.30 &  & 1079.53 & 13.0 & 33.28 & 0.07 &  & 1080.11 & 13.0 & 33.28 & 0.13 &  & \textbf{1078.75} & 13 \\ 
UK100\_04 &  & 1106.48 & 14 & 149.70 & 2.90 &  & 1076.46 & 14.0 & 35.79 & 0.11 &  & 1077.42 & 14.0 & 35.79 & 0.20 &  & \textbf{1075.29} & 14 \\ 
UK100\_05 &  & 1043.41 & 14 & 159.00 & 1.41 &  & 1032.64 & 14.4 & 33.63 & 0.37 &  & 1036.51 & 14.4 & 33.63 & 0.74 &  & \textbf{1028.86} & 14 \\ 
UK100\_06 &  & 1213.61 & 14 & 133.80 & 1.70 &  & 1193.86 & 14.0 & 29.37 & 0.04 &  & 1195.60 & 14.0 & 29.37 & 0.19 &  & \textbf{1193.38} & 14 \\ 
UK100\_07 &  & 1060.08 & 12 & 102.60 & 1.44 &  & 1046.92 & 12.0 & 28.63 & 0.18 &  & 1047.66 & 12.0 & 28.63 & 0.25 &  & \textbf{1045.02} & 12 \\ 
UK100\_08 &  & 1106.78 & 13 & 209.50 & 1.55 &  & 1091.27 & 12.6 & 26.63 & 0.13 &  & 1092.70 & 12.6 & 26.63 & 0.26 &  & \textbf{1089.84} & 12 \\ 
UK100\_09 &  & 1015.46 & 13 & 154.00 & 2.74 &  & 989.66 & 13.0 & 30.47 & 0.13 &  & 991.18 & 13.0 & 30.47 & 0.28 &  & \textbf{988.41} & 13 \\ 
UK100\_10 &  & 1076.56 & 12 & 199.00 & 1.57 &  & 1061.42 & 12.0 & 29.73 & 0.14 &  & 1061.45 & 12.0 & 29.73 & 0.14 &  & \textbf{1059.95} & 12 \\ 
UK100\_11 &  & 1210.25 & 15 & 107.10 & 1.15 &  & 1198.08 & 14.0 & 36.26 & 0.13 &  & 1202.36 & 14.1 & 36.26 & 0.49 &  & \textbf{1196.50} & 14 \\ 
UK100\_12 &  & 1053.02 & 12 & 206.40 & 2.50 &  & 1028.95 & 12.0 & 31.91 & 0.15 &  & 1029.86 & 12.0 & 31.91 & 0.24 &  & \textbf{1027.38} & 12 \\ 
UK100\_13 &  & 1154.83 & 13 & 87.90 & 2.01 &  & 1132.72 & 13.0 & 27.46 & 0.06 &  & 1134.15 & 13.0 & 27.46 & 0.19 &  & \textbf{1132.03} & 13 \\ 
UK100\_14 &  & 1264.50 & 14 & 91.80 & 1.76 &  & 1242.68 & 14.0 & 31.45 & 0.00 &  & 1243.67 & 14.0 & 31.45 & 0.08 &  & \textbf{1242.68} & 14 \\ 
UK100\_15 &  & 1315.50 & 15 & 110.90 & 1.18 &  & 1302.19 & 15.0 & 36.24 & 0.16 &  & 1303.81 & 15.0 & 36.24 & 0.28 &  & \textbf{1300.13} & 15 \\ 
UK100\_16 &  & 1005.03 & 12 & 254.70 & 2.36 &  & 982.77 & 12.0 & 28.14 & 0.09 &  & 984.52 & 12.0 & 28.14 & 0.27 &  & \textbf{981.86} & 12 \\ 
UK100\_17 &  & 1284.81 & 15 & 152.80 & 2.12 &  & 1259.07 & 15.0 & 38.88 & 0.07 &  & 1259.27 & 15.0 & 38.88 & 0.09 &  & \textbf{1258.16} & 15 \\ 
UK100\_18 &  & 1106.00 & 13 & 92.60 & 3.04 &  & 1079.79 & 12.9 & 33.15 & 0.60 &  & 1081.40 & 12.9 & 33.15 & 0.75 &  & \textbf{1073.38} & 12 \\ 
UK100\_19 &  & 1044.71 & 13 & 91.00 & 2.83 &  & 1017.22 & 13.0 & 30.67 & 0.13 &  & 1018.71 & 13.0 & 30.67 & 0.27 &  & \textbf{1015.95} & 13 \\ 
UK100\_20 &  & 1263.06 & 14 & 204.40 & 1.86 &  & 1241.72 & 14.0 & 30.05 & 0.14 &  & 1244.41 & 14.0 & 30.05 & 0.36 &  & \textbf{1240.00} & 14 \\ \hline
\textbf{Avg.} &  &  &  & \textbf{145.27} & \textbf{1.99} &  &  &  & \textbf{31.98} & \textbf{0.15} &  &  &  & \textbf{31.98} & \textbf{0.28} &  &  &  \\ \hline
UK100\_01-B &  & -- & -- & -- & -- &  & 1593.01 & 14.0 & 83.29 & 0.11 &  & 1594.84 & 14.0 & 83.29 & 0.23 &  & \textbf{1591.20} & 14 \\ 
UK100\_02-B &  & -- & -- & -- & -- &  & 1602.66 & 13.0 & 96.50 & 0.19 &  & 1603.01 & 13.0 & 96.50 & 0.22 &  & \textbf{1599.56} & 13 \\ 
UK100\_03-B &  & -- & -- & -- & -- &  & 1509.47 & 13.0 & 229.98 & 0.60 &  & 1506.67 & 13.0 & 229.98 & 0.42 &  & \textbf{1500.40} & 13 \\ 
UK100\_04-B &  & -- & -- & -- & -- &  & 1472.97 & 14.0 & 117.36 & 0.03 &  & 1473.84 & 14.0 & 117.36 & 0.09 &  & \textbf{1472.49} & 14 \\ 
UK100\_05-B &  & -- & -- & -- & -- &  & 1490.68 & 14.0 & 108.81 & 0.13 &  & 1493.42 & 14.2 & 108.81 & 0.32 &  & \textbf{1488.73} & 15 \\ 
UK100\_06-B &  & -- & -- & -- & -- &  & 1648.66 & 14.0 & 56.75 & 0.18 &  & 1649.37 & 14.0 & 56.75 & 0.22 &  & \textbf{1645.77} & 14 \\ 
UK100\_07-B &  & -- & -- & -- & -- &  & 1509.65 & 12.0 & 80.10 & 0.11 &  & 1510.73 & 12.0 & 80.10 & 0.18 &  & \textbf{1508.05} & 12 \\ 
UK100\_08-B &  & -- & -- & -- & -- &  & 1468.61 & 12.0 & 110.27 & 0.14 &  & 1477.29 & 12.5 & 110.27 & 0.73 &  & \textbf{1466.62} & 12 \\ 
UK100\_09-B &  & -- & -- & -- & -- &  & 1380.06 & 13.0 & 60.76 & 0.18 &  & 1380.30 & 13.0 & 60.76 & 0.19 &  & \textbf{1377.64} & 13 \\ 
UK100\_10-B &  & -- & -- & -- & -- &  & 1479.30 & 12.0 & 73.18 & 0.14 &  & 1479.82 & 12.0 & 73.18 & 0.17 &  & \textbf{1477.25} & 12 \\ 
UK100\_11-B &  & -- & -- & -- & -- &  & 1623.07 & 14.0 & 71.22 & 0.26 &  & 1630.02 & 14.0 & 71.22 & 0.68 &  & \textbf{1618.94} & 14 \\ 
UK100\_12-B &  & -- & -- & -- & -- &  & 1362.53 & 12.0 & 91.03 & 0.03 &  & 1366.52 & 12.1 & 91.03 & 0.32 &  & \textbf{1362.14} & 12 \\ 
UK100\_13-B &  & -- & -- & -- & -- &  & 1606.86 & 13.0 & 68.06 & 0.05 &  & 1606.47 & 13.0 & 68.06 & 0.03 &  & \textbf{1605.99} & 13 \\ 
UK100\_14-B &  & -- & -- & -- & -- &  & 1690.26 & 14.0 & 76.47 & 0.00 &  & 1690.25 & 14.0 & 76.47 & 0.00 &  & \textbf{1690.25} & 14 \\ 
UK100\_15-B &  & -- & -- & -- & -- &  & 1736.64 & 15.0 & 56.36 & 0.10 &  & 1735.18 & 15.0 & 56.36 & 0.01 &  & \textbf{1734.92} & 15 \\ 
UK100\_16-B &  & -- & -- & -- & -- &  & 1382.00 & 12.0 & 78.23 & 0.07 &  & 1386.92 & 12.1 & 78.23 & 0.43 &  & \textbf{1381.03} & 13 \\ 
UK100\_17-B &  & -- & -- & -- & -- &  & 1679.13 & 15.0 & 118.92 & 0.06 &  & 1678.78 & 15.0 & 118.92 & 0.04 &  & \textbf{1678.04} & 15 \\ 
UK100\_18-B &  & -- & -- & -- & -- &  & 1505.79 & 12.7 & 128.55 & 0.68 &  & 1508.96 & 12.9 & 128.55 & 0.89 &  & \textbf{1495.60} & 13 \\ 
UK100\_19-B &  & -- & -- & -- & -- &  & 1401.56 & 13.0 & 86.72 & 0.09 &  & 1401.65 & 13.0 & 86.72 & 0.10 &  & \textbf{1400.28} & 13 \\ 
UK100\_20-B &  & -- & -- & -- & -- &  & 1629.51 & 14.0 & 47.20 & 0.04 &  & 1629.15 & 14.0 & 47.20 & 0.02 &  & \textbf{1628.89} & 14 \\ \hline
\textbf{Avg.} &  &  &  & \textbf{--} & \textbf{--} &  &  &  & \textbf{91.99} & \textbf{0.16} &  &  &  & \textbf{91.99} & \textbf{0.26} &  &  &  \\ \hline
UK100\_01-C &  & -- & -- & -- & -- &  & 1489.40 & 14.0 & 49.58 & 0.21 &  & 1492.63 & 14.0 & 49.58 & 0.42 &  & \textbf{1486.34} & 14 \\ 
UK100\_02-C &  & -- & -- & -- & -- &  & 1432.18 & 13.0 & 56.86 & 0.04 &  & 1432.84 & 13.0 & 56.86 & 0.09 &  & \textbf{1431.55} & 13 \\ 
UK100\_03-C &  & -- & -- & -- & -- &  & 1325.41 & 13.0 & 65.95 & 0.20 &  & 1326.02 & 13.0 & 65.95 & 0.25 &  & \textbf{1322.73} & 13 \\ 
UK100\_04-C &  & -- & -- & -- & -- &  & 1382.10 & 14.0 & 80.26 & 0.32 &  & 1382.50 & 14.0 & 80.26 & 0.35 &  & \textbf{1377.73} & 14 \\ 
UK100\_05-C &  & -- & -- & -- & -- &  & 1308.92 & 14.0 & 55.27 & 0.22 &  & 1310.80 & 14.0 & 55.27 & 0.36 &  & \textbf{1306.04} & 14 \\ 
UK100\_06-C &  & -- & -- & -- & -- &  & 1486.69 & 14.0 & 46.01 & 0.05 &  & 1487.86 & 14.0 & 46.01 & 0.13 &  & \textbf{1485.99} & 14 \\ 
UK100\_07-C &  & -- & -- & -- & -- &  & 1333.24 & 12.0 & 74.42 & 0.12 &  & 1334.78 & 12.0 & 74.42 & 0.23 &  & \textbf{1331.67} & 12 \\ 
UK100\_08-C &  & -- & -- & -- & -- &  & 1380.30 & 12.3 & 62.02 & 0.52 &  & 1380.89 & 12.0 & 62.02 & 0.56 &  & \textbf{1373.20} & 13 \\ 
UK100\_09-C &  & -- & -- & -- & -- &  & 1275.09 & 12.9 & 52.63 & 0.33 &  & 1275.60 & 12.8 & 52.63 & 0.37 &  & \textbf{1270.84} & 13 \\ 
UK100\_10-C &  & -- & -- & -- & -- &  & 1334.30 & 12.0 & 56.29 & 0.33 &  & 1335.22 & 12.0 & 56.29 & 0.40 &  & \textbf{1329.95} & 12 \\ 
UK100\_11-C &  & -- & -- & -- & -- &  & 1499.15 & 14.0 & 44.47 & 0.00 &  & 1500.60 & 14.0 & 44.47 & 0.10 &  & \textbf{1499.15} & 14 \\ 
UK100\_12-C &  & -- & -- & -- & -- &  & 1235.54 & 12.0 & 52.03 & 0.18 &  & 1237.53 & 12.0 & 52.03 & 0.34 &  & \textbf{1233.28} & 12 \\ 
UK100\_13-C &  & -- & -- & -- & -- &  & 1443.88 & 13.0 & 56.31 & 0.09 &  & 1443.21 & 13.0 & 56.31 & 0.04 &  & \textbf{1442.65} & 13 \\ 
UK100\_14-C &  & -- & -- & -- & -- &  & 1554.46 & 14.0 & 50.09 & 0.10 &  & 1555.29 & 14.0 & 50.09 & 0.16 &  & \textbf{1552.85} & 14 \\ 
UK100\_15-C &  & -- & -- & -- & -- &  & 1626.99 & 15.0 & 64.80 & 0.08 &  & 1627.87 & 15.0 & 64.80 & 0.14 &  & \textbf{1625.66} & 15 \\ 
UK100\_16-C &  & -- & -- & -- & -- &  & 1218.92 & 12.0 & 47.35 & 0.17 &  & 1219.92 & 12.0 & 47.35 & 0.25 &  & \textbf{1216.84} & 12 \\ 
UK100\_17-C &  & -- & -- & -- & -- &  & 1558.93 & 15.0 & 109.41 & 0.35 &  & 1563.23 & 15.0 & 109.41 & 0.63 &  & \textbf{1553.50} & 15 \\ 
UK100\_18-C &  & -- & -- & -- & -- &  & 1331.53 & 12.5 & 72.65 & 0.78 &  & 1336.22 & 12.7 & 72.65 & 1.14 &  & \textbf{1321.19} & 13 \\ 
UK100\_19-C &  & -- & -- & -- & -- &  & 1274.31 & 13.0 & 83.30 & 0.11 &  & 1275.42 & 13.0 & 83.30 & 0.19 &  & \textbf{1272.96} & 13 \\ 
UK100\_20-C &  & -- & -- & -- & -- &  & 1540.42 & 14.0 & 61.09 & 0.21 &  & 1540.99 & 14.0 & 61.09 & 0.25 &  & \textbf{1537.13} & 14 \\ \hline
\textbf{Avg.} &  &  &  & \textbf{--} & \textbf{--} &  &  &  & \textbf{62.04} & \textbf{0.22} &  &  &  & \textbf{62.04} & \textbf{0.32} &  &  &  \\ \hline
\multicolumn{10}{l}{\scriptsize{$^{*}$ 3 GHz CPU with 1 GB of RAM}}
\end{tabular}
\label{resultsPRP100}
}
\end{table}
\begin{table}[htbp]
\centering
\renewcommand{\arraystretch}{0.92}
\caption{Results for the PRP 200-customer instances}
\small
\scalebox{0.9}
{
\setlength{\tabcolsep}{1mm}
\begin{tabular}{ccccccc@{\hspace*{0.4cm}}ccccc@{\hspace*{0.4cm}}ccccc@{\hspace*{0.4cm}}cc}
\hline
\textbf{} & \textbf{} & \multicolumn{ 4}{c}{\textbf{ALNS}} &  & \multicolumn{4}{c}{\textbf{Avg. ILS-SP-SOA-Dyn}} &  & \multicolumn{ 4}{c}{\textbf{Avg. ILS-SP-SOA-Stat}} &  & \multicolumn{ 2}{c}{\textbf{BKS}} \\ \cline{ 3- 6} \cline{ 8- 11}\cline{ 13- 16}\cline{ 18- 19}
\textbf{Instance} & \textbf{} & \textbf{} & \textbf{} & \textbf{CPU} & \textbf{Gap} & \textbf{} & \textbf{} & \textbf{} & \textbf{CPU} & \textbf{Gap} & \textbf{} & \textbf{} & \textbf{} & \textbf{CPU} & \textbf{Gap} & \textbf{} & \textbf{} & \textbf{} \\
\multicolumn{ 1}{c}{} & \textbf{} & \textbf{\up{Cost}} & \textbf{\up{$|\mathbf{R}|$}} & \textbf{T(s)$^{*}$} & \textbf{(\%)} & \textbf{} & \textbf{\up{Cost}} & \textbf{\up{$|\mathbf{R}|$}} & \textbf{T(s)} & \textbf{(\%)} & \textbf{} & \textbf{\up{Cost}} & \textbf{\up{$|\mathbf{R}|$}} & \textbf{T(s)} & \textbf{(\%)} & \textbf{} & \textbf{\up{Cost}} & \textbf{\up{$|\mathbf{R}|$}} \\ \hline
UK200\_01 &  & 2111.70 & 28 & 724.40 & 4.86 &  & 2029.75 & 27.9 & 305.52 & 0.79 &  & 2034.00 & 27.8 & 305.52 & 1.00 &  & \textbf{2013.84} & 27 \\ 
UK200\_02 &  & 1988.64 & 24 & 1020.90 & 4.39 &  & 1917.08 & 24.0 & 297.58 & 0.63 &  & 1916.00 & 24.0 & 297.58 & 0.57 &  & \textbf{1905.08} & 24 \\ 
UK200\_03 &  & 2017.63 & 27 & 404.10 & 3.55 &  & 1956.62 & 27.0 & 275.43 & 0.42 &  & 1955.00 & 27.0 & 275.43 & 0.33 &  & \textbf{1948.50} & 27 \\ 
UK200\_04 &  & 1934.13 & 26 & 411.70 & 4.37 &  & 1865.68 & 26.1 & 312.24 & 0.68 &  & 1866.00 & 26.0 & 312.24 & 0.69 &  & \textbf{1853.16} & 26 \\ 
UK200\_05 &  & 2182.91 & 27 & 926.40 & 3.88 &  & 2108.80 & 27.0 & 288.20 & 0.36 &  & 2111.00 & 27.0 & 288.20 & 0.46 &  & \textbf{2101.34} & 27 \\ 
UK200\_06 &  & 1883.22 & 27 & 450.30 & 3.94 &  & 1817.77 & 27.0 & 291.23 & 0.33 &  & 1822.00 & 26.9 & 291.23 & 0.56 &  & \textbf{1811.85} & 27 \\ 
UK200\_07 &  & 2021.95 & 27 & 943.40 & 4.96 &  & 1939.79 & 27.0 & 303.20 & 0.69 &  & 1941.00 & 27.0 & 303.20 & 0.75 &  & \textbf{1926.47} & 27 \\ 
UK200\_08 &  & 2116.76 & 27 & 430.60 & 4.03 &  & 2043.37 & 27.0 & 294.33 & 0.43 &  & 2046.00 & 27.0 & 294.33 & 0.56 &  & \textbf{2034.68} & 27 \\ 
UK200\_09 &  & 1894.18 & 25 & 553.10 & 6.04 &  & 1793.64 & 25.1 & 271.33 & 0.41 &  & 1796.00 & 25.3 & 271.33 & 0.55 &  & \textbf{1786.24} & 25 \\ 
UK200\_10 &  & 2199.95 & 28 & 500.00 & 3.55 &  & 2134.94 & 27.1 & 273.15 & 0.49 &  & 2135.00 & 27.1 & 273.15 & 0.50 &  & \textbf{2124.44} & 27 \\ 
UK200\_11 &  & 1941.19 & 27 & 842.80 & 4.75 &  & 1861.65 & 27.0 & 292.80 & 0.46 &  & 1861.00 & 27.0 & 292.80 & 0.43 &  & \textbf{1853.12} & 27 \\ 
UK200\_12 &  & 2105.14 & 25 & 711.00 & 2.42 &  & 2067.11 & 25.0 & 254.49 & 0.57 &  & 2069.00 & 25.0 & 254.49 & 0.67 &  & \textbf{2055.32} & 25 \\ 
UK200\_13 &  & 2141.26 & 25 & 444.60 & 4.42 &  & 2064.18 & 25.1 & 304.74 & 0.66 &  & 2066.00 & 25.0 & 304.74 & 0.75 &  & \textbf{2050.70} & 25 \\ 
UK200\_14 &  & 2011.35 & 27 & 450.60 & 3.57 &  & 1950.56 & 27.0 & 297.60 & 0.44 &  & 1952.00 & 27.0 & 297.60 & 0.51 &  & \textbf{1942.10} & 27 \\ 
UK200\_15 &  & 2110.86 & 25 & 542.00 & 5.22 &  & 2018.80 & 25.7 & 288.89 & 0.63 &  & 2020.00 & 25.7 & 288.89 & 0.69 &  & \textbf{2006.21} & 25 \\ 
UK200\_16 &  & 2075.83 & 27 & 455.60 & 4.76 &  & 1991.29 & 27.0 & 276.39 & 0.49 &  & 1994.00 & 27.0 & 276.39 & 0.63 &  & \textbf{1981.57} & 27 \\ 
UK200\_17 &  & 2218.28 & 26 & 409.20 & 4.55 &  & 2126.25 & 26.0 & 332.38 & 0.21 &  & 2126.00 & 26.0 & 332.38 & 0.20 &  & \textbf{2121.77} & 26 \\ 
UK200\_18 &  & 2004.68 & 27 & 788.90 & 3.12 &  & 1961.48 & 27.0 & 316.89 & 0.90 &  & 1961.00 & 26.8 & 316.89 & 0.87 &  & \textbf{1944.03} & 26 \\ 
UK200\_19 &  & 1844.90 & 25 & 973.90 & 4.70 &  & 1770.10 & 25.0 & 306.28 & 0.45 &  & 1768.00 & 25.0 & 306.28 & 0.33 &  & \textbf{1762.14} & 25 \\ 
UK200\_20 &  & 2150.57 & 27 & 531.10 & 3.71 &  & 2089.33 & 26.2 & 350.92 & 0.75 &  & 2088.00 & 26.3 & 350.92 & 0.69 &  & \textbf{2073.68} & 26 \\ \hline
\textbf{Avg.} &  &  &  & \textbf{625.73} & \textbf{4.24} &  &  &  & \textbf{296.68} & \textbf{0.54} &  &  &  & \textbf{296.68} & \textbf{0.59} &  &  &  \\ \hline
UK200\_01-B &  & -- & -- & -- & -- &  & 2744.22 & 27.8 & 1583.51 & 0.69 &  & 2742.42 & 28.0 & 1583.51 & 0.63 &  & \textbf{2725.35} & 28 \\ 
UK200\_02-B &  & -- & -- & -- & -- &  & 2581.36 & 24.0 & 853.37 & 0.40 &  & 2585.83 & 24.0 & 853.37 & 0.58 &  & \textbf{2570.95} & 25 \\ 
UK200\_03-B &  & -- & -- & -- & -- &  & 2764.51 & 27.0 & 1344.78 & 0.57 &  & 2764.05 & 27.0 & 1344.78 & 0.55 &  & \textbf{2748.82} & 28 \\ 
UK200\_04-B &  & -- & -- & -- & -- &  & 2602.43 & 26.0 & 1066.47 & 0.95 &  & 2603.11 & 26.0 & 1066.47 & 0.97 &  & \textbf{2578.03} & 26 \\ 
UK200\_05-B &  & -- & -- & -- & -- &  & 2820.23 & 26.9 & 1079.18 & 0.74 &  & 2819.43 & 26.8 & 1079.18 & 0.71 &  & \textbf{2799.43} & 27 \\ 
UK200\_06-B &  & -- & -- & -- & -- &  & 2584.88 & 26.1 & 1188.31 & 0.71 &  & 2582.40 & 26.1 & 1188.31 & 0.61 &  & \textbf{2566.74} & 27 \\ 
UK200\_07-B &  & -- & -- & -- & -- &  & 2668.12 & 27.0 & 1260.38 & 0.64 &  & 2666.86 & 27.0 & 1260.38 & 0.60 &  & \textbf{2651.03} & 27 \\ 
UK200\_08-B &  & -- & -- & -- & -- &  & 2774.81 & 27.0 & 1882.30 & 0.74 &  & 2779.02 & 27.0 & 1882.30 & 0.89 &  & \textbf{2754.46} & 27 \\ 
UK200\_09-B &  & -- & -- & -- & -- &  & 2506.86 & 25.0 & 812.10 & 0.76 &  & 2501.42 & 25.0 & 812.10 & 0.54 &  & \textbf{2488.01} & 25 \\ 
UK200\_10-B &  & -- & -- & -- & -- &  & 2895.74 & 27.1 & 774.30 & 0.43 &  & 2895.69 & 27.0 & 774.30 & 0.42 &  & \textbf{2883.48} & 28 \\ 
UK200\_11-B &  & -- & -- & -- & -- &  & 2609.10 & 27.0 & 824.49 & 1.09 &  & 2602.88 & 27.0 & 824.49 & 0.85 &  & \textbf{2580.95} & 27 \\ 
UK200\_12-B &  & -- & -- & -- & -- &  & 2768.39 & 25.0 & 1494.45 & 0.54 &  & 2765.16 & 25.0 & 1494.45 & 0.43 &  & \textbf{2753.44} & 25 \\ 
UK200\_13-B &  & -- & -- & -- & -- &  & 2747.43 & 25.0 & 1033.02 & 0.66 &  & 2743.10 & 25.0 & 1033.02 & 0.50 &  & \textbf{2729.44} & 26 \\ 
UK200\_14-B &  & -- & -- & -- & -- &  & 2695.51 & 27.0 & 606.18 & 0.48 &  & 2697.47 & 27.0 & 606.18 & 0.55 &  & \textbf{2682.63} & 27 \\ 
UK200\_15-B &  & -- & -- & -- & -- &  & 2788.26 & 25.3 & 1228.65 & 0.82 &  & 2791.44 & 25.6 & 1228.65 & 0.93 &  & \textbf{2765.68} & 26 \\ 
UK200\_16-B &  & -- & -- & -- & -- &  & 2745.83 & 27.0 & 1285.35 & 0.82 &  & 2746.78 & 27.0 & 1285.35 & 0.85 &  & \textbf{2723.54} & 27 \\ 
UK200\_17-B &  & -- & -- & -- & -- &  & 2825.82 & 26.0 & 1329.93 & 0.38 &  & 2823.86 & 26.0 & 1329.93 & 0.31 &  & \textbf{2815.16} & 26 \\ 
UK200\_18-B &  & -- & -- & -- & -- &  & 2733.79 & 26.6 & 961.21 & 0.93 &  & 2732.07 & 26.3 & 961.21 & 0.87 &  & \textbf{2708.62} & 27 \\ 
UK200\_19-B &  & -- & -- & -- & -- &  & 2456.58 & 25.0 & 1083.72 & 0.43 &  & 2451.46 & 25.0 & 1083.72 & 0.22 &  & \textbf{2446.17} & 25 \\ 
UK200\_20-B &  & -- & -- & -- & -- &  & 2825.74 & 26.0 & 1555.97 & 0.70 &  & 2819.87 & 26.0 & 1555.97 & 0.49 &  & \textbf{2806.02} & 27 \\ \hline
\textbf{Avg.} &  &  &  & \textbf{--} & \textbf{--} &  &  &  & \textbf{1162.38} & \textbf{0.67} &  &  &  & \textbf{1162.38} & \textbf{0.63} &  &  &  \\ \hline
UK200\_01-C &  & -- & -- & -- & -- &  & 2563.48 & 27.3 & 626.96 & 0.87 &  & 2566.65 & 27.6 & 626.96 & 1.00 &  & \textbf{2541.36} & 28 \\ 
UK200\_02-C &  & -- & -- & -- & -- &  & 2427.91 & 24.0 & 498.31 & 0.82 &  & 2430.31 & 24.0 & 498.31 & 0.92 &  & \textbf{2408.27} & 24 \\ 
UK200\_03-C &  & -- & -- & -- & -- &  & 2506.35 & 27.0 & 360.02 & 0.73 &  & 2506.63 & 27.0 & 360.02 & 0.74 &  & \textbf{2488.24} & 27 \\ 
UK200\_04-C &  & -- & -- & -- & -- &  & 2402.22 & 26.0 & 446.52 & 1.39 &  & 2396.98 & 26.0 & 446.52 & 1.17 &  & \textbf{2369.21} & 26 \\ 
UK200\_05-C &  & -- & -- & -- & -- &  & 2603.07 & 27.0 & 549.66 & 1.13 &  & 2601.74 & 27.0 & 549.66 & 1.08 &  & \textbf{2573.91} & 27 \\ 
UK200\_06-C &  & -- & -- & -- & -- &  & 2350.51 & 26.0 & 589.35 & 0.87 &  & 2350.39 & 26.1 & 589.35 & 0.86 &  & \textbf{2330.26} & 27 \\ 
UK200\_07-C &  & -- & -- & -- & -- &  & 2461.99 & 27.0 & 637.76 & 0.89 &  & 2464.48 & 27.0 & 637.76 & 0.99 &  & \textbf{2440.32} & 27 \\ 
UK200\_08-C &  & -- & -- & -- & -- &  & 2514.24 & 27.0 & 748.53 & 0.81 &  & 2513.47 & 27.0 & 748.53 & 0.78 &  & \textbf{2493.97} & 27 \\ 
UK200\_09-C &  & -- & -- & -- & -- &  & 2321.51 & 25.0 & 448.07 & 1.00 &  & 2321.77 & 25.0 & 448.07 & 1.01 &  & \textbf{2298.45} & 25 \\ 
UK200\_10-C &  & -- & -- & -- & -- &  & 2591.84 & 27.0 & 406.03 & 0.67 &  & 2591.10 & 27.0 & 406.03 & 0.64 &  & \textbf{2574.70} & 27 \\ 
UK200\_11-C &  & -- & -- & -- & -- &  & 2409.41 & 27.0 & 469.44 & 0.77 &  & 2412.57 & 27.0 & 469.44 & 0.91 &  & \textbf{2390.89} & 27 \\ 
UK200\_12-C &  & -- & -- & -- & -- &  & 2556.28 & 25.0 & 521.46 & 0.49 &  & 2556.63 & 25.0 & 521.46 & 0.51 &  & \textbf{2543.77} & 25 \\ 
UK200\_13-C &  & -- & -- & -- & -- &  & 2517.62 & 25.0 & 685.04 & 1.09 &  & 2513.41 & 25.0 & 685.04 & 0.92 &  & \textbf{2490.52} & 26 \\ 
UK200\_14-C &  & -- & -- & -- & -- &  & 2495.24 & 27.0 & 447.67 & 0.90 &  & 2492.50 & 27.0 & 447.67 & 0.79 &  & \textbf{2472.98} & 27 \\ 
UK200\_15-C &  & -- & -- & -- & -- &  & 2541.33 & 25.2 & 707.91 & 1.20 &  & 2540.70 & 25.4 & 707.91 & 1.17 &  & \textbf{2511.23} & 26 \\ 
UK200\_16-C &  & -- & -- & -- & -- &  & 2530.32 & 27.0 & 473.10 & 1.25 &  & 2529.27 & 27.0 & 473.10 & 1.21 &  & \textbf{2499.09} & 27 \\ 
UK200\_17-C &  & -- & -- & -- & -- &  & 2597.37 & 26.0 & 416.65 & 0.68 &  & 2600.91 & 26.0 & 416.65 & 0.82 &  & \textbf{2579.84} & 26 \\ 
UK200\_18-C &  & -- & -- & -- & -- &  & 2460.29 & 26.4 & 641.90 & 1.34 &  & 2459.66 & 26.5 & 641.90 & 1.31 &  & \textbf{2427.76} & 27 \\ 
UK200\_19-C &  & -- & -- & -- & -- &  & 2275.02 & 25.0 & 443.44 & 0.84 &  & 2275.04 & 25.0 & 443.44 & 0.84 &  & \textbf{2255.98} & 25 \\ 
UK200\_20-C &  & -- & -- & -- & -- &  & 2551.52 & 26.0 & 553.67 & 0.84 &  & 2549.18 & 26.0 & 553.67 & 0.75 &  & \textbf{2530.16} & 26 \\ \hline
\textbf{Avg.} &  &  &  & \textbf{--} & \textbf{--} &  &  &  & \textbf{533.57} & \textbf{0.93} &  &  &  & \textbf{533.57} & \textbf{0.92} &  &  &  \\ \hline
\multicolumn{10}{l}{\scriptsize{$^{*}$ 3 GHz CPU with 1 GB of RAM}}
\end{tabular}
\label{resultsPRP200}
}
\end{table}


\end{document}